\documentclass[twocolumn,tighten]{aastex63}

\usepackage{amsmath}
\usepackage{comment}
\usepackage{tabularx}
\usepackage{algorithm2e}
\usepackage{float}
\usepackage{fullwidth}
\usepackage{nccmath}
\usepackage{mathtools}
\usepackage{multirow}
\usepackage{scrextend}
\usepackage{lineno}
\usepackage{rotating}

\definecolor{yscol}{rgb}{0.8, 0.6, 1}

\newcommand{\msun}{\,{\rm M}_\odot}
\newcommand{\erg}{\,{\rm erg}}
\newcommand{\pc}{\,{\rm pc}}
\newcommand{\kpc}{\,{\rm kpc}}
\newcommand{\km}{\,{\rm km}}
\newcommand{\s}{\,{\rm s}}
\newcommand{\g}{\,{\rm g}}
\newcommand{\cm}{\,{\rm cm}}
\newcommand{\yr}{\,{\rm yr}}
\newcommand{\Myr}{\,{\rm Myr}}
\newcommand{\Gyr}{\,{\rm Gyr}}

\newcommand{\myr}{\,{\rm Myr}}

\newcommand{\newenzo}{{\tt Enzo-{\tt Abyss}}}
\newcommand{\oldenzo}{{\tt Enzo-N}}

\newcommand{\enzo}{{\tt Enzo}}
\newcommand{\nbody}{{\tt Abyss}}
\definecolor{yscol}{rgb}{0.8, 0.6, 1}

\def\bs#1{{\boldsymbol{#1}}}

\newcommand{\tr}[1]{{#1}}

\newlength{\Oldarrayrulewidth}

\makeatletter
\newcommand{\thickhline}{%
    \noalign {\ifnum 0=`}\fi \hrule height 1pt
    \futurelet \reserved@a \@xhline
}
\newcommand{\thichline}{%
    \noalign {\ifnum 0=`}\fi \hrule height 0.8pt
    \futurelet \reserved@a \@xhline
}
\newcolumntype{"}{@{\hskip\tabcolsep\vrule width 0.8pt\hskip\tabcolsep}} 
\makeatother

\received{June 1, 2019}
\revised{\today}
\accepted{January 10, 2025}
\submitjournal{ApJ}

\graphicspath{{./}{figures/}}

\begin{document}





\title{Evolution of Nuclear Star Cluster in Dwarf Galaxy through Mergers and In-Situ Star Formation}

\correspondingauthor{Yongseok Jo}
\email{yj2812@columbia.edu}
\author[0000-0003-3977-1761]{Yongseok Jo}
\affiliation{Columbia Astrophysics Laboratory, Columbia University, New York, NY 10027, USA}
\affiliation{Center for Computational Astrophysics, Flatiron Institute, 162 5th Avenue, New York, NY, 10010, USA}

\author[0000-0002-9144-1383]{Minyong Jung}
\affiliation{Center for Theoretical Physics, Department of Physics and Astronomy, Seoul National University, Seoul 08826, Korea}

\author{Greg L. Bryan}
\affiliation{Department of Astronomy, Columbia University, New York, NY 10027, USA}

\author{Seoyoung Kim}
\affiliation{Department of Biophysics, University of Wisconsin-Madison, Madison, WI 53703, USA}

\author[0000-0003-4464-1160]{Ji-hoon Kim}
\affiliation{Center for Theoretical Physics, Department of Physics and Astronomy, Seoul National University, Seoul 08826, Korea}
\affiliation{Institute for Data Innovation in Science, Seoul National University, Seoul 08826, Korea}
\affiliation{Seoul National University Astronomy Research Center, Seoul 08826, Korea}

\author{Ahram Lee}
\affiliation{Synopsys Korea Inc., Gyeonggi-do 13494, Korea}



\begin{abstract}
Nuclear Star Clusters (NSCs) are dense stellar systems located at the centers of galaxies. 
Employing \newenzo{}, which integrates hydrodynamics with a direct N-body solver, 
we introduce a simulation capable of resolving the evolution of NSCs within a live galaxy. 
This includes live dark matter, gaseous dynamics, star formation and feedback, collisional dynamics for star clusters. 
The evolution of NSCs is typically shaped by two main processes: mergers of star clusters and in-situ star formation. 
Our simulation enables investigation of the contributions of these mechanisms to the growth of NSCs.
This work focuses on the impact of stellar physics and gas content on the growth of NSCs within a dwarf galaxy.
To this end, we carry out four simulations, a fiducial simulation, one without supernova feedback, one with low star formation efficiency, and one with higher galactic gas content.
This study shows a likelihood that both mergers and in-situ star formation contribute to NSC evolution comparably.
In addition, mergers result in disruption of dense gas clumps within star clusters, indicating that in-situ star formation is suppressed when mergers occur.
However, the limitations---such as the lack of individual star physics and limited spatial/particle mass resolution---hinder drawing a definite conclusion.
Nevertheless, with further development, our simulations will serve as a cornerstone that untangles the complex interplay between mergers and in-situ star formation in shaping the structure and mass of NSCs, thereby providing insights into their formation and evolution.
\end{abstract}


\section{Introduction}
\label{sec:introduction}


Nuclear star clusters (NSCs) are among the densest stellar systems in the universe, residing at the centers of galaxies.
Their high luminosities made them identifiable even in early surveys \citep{becklin1968ApJ...151..145B, becklin1975ApJ...200L..71B, binggeli1987AJ.....94..251B, kormendy1989ARA&A..27..235K, sarajendini1996AAS...188.0306S, phillips1996AJ....111.1566P, carollo2002AJ....123..159C, hughes2005AJ....130...73H}, and subsequent observations have shown that they are ubiquitous across galaxy types: dwarf galaxies \citep{reaves1983ApJS...53..375R, caldwell1983AJ.....88..804C, lotz2001ApJ...552..572L, georgiev2009MNRAS.396.1075G, onrdenes2018ApJ...860....4O, sanchez2019ApJ...878...18S}, early-type galaxies \citep{cote2006ApJS..165...57C, denbrok2014MNRAS.445.2385D, turner2012ApJS..203....5T, baldassare2014ApJ...791..133B}, late-type galaxies \citep{boker2002AJ....123.1389B, walcher2006ApJ...649..692W, seth2006AJ....132.2539S}, and low surface brightness galaxies \citep{khim2024AJ....168...45K}.
The frequency of nucleation and the NSC-to-host mass ratio both depend on the stellar mass and morphological type of the host \citep{Neumayer2020review}, giving rise to scaling relations such as the bulge--nuclei relation \citep{balcells2003ApJ...582L..79B, balcells2007ApJ...665.1084B} and the nucleated fraction--luminosity relation \citep{ferrarese2006ApJ...644L..21F}.
Recent JWST observations have further expanded our view of NSCs across a wide range of redshifts, probing both their stellar ages and gas properties \citep{fahrion2024A&A...687A..83F, pfeffer2024arXiv241007498P, doan2024arXiv240804774D, buiten2024ApJ...966..166B}.

The compact nature of NSCs, coupled with limited resolution, typically hinders detailed characterization.
Nevertheless, it is established that their typical masses span $10^4$--$10^8\msun$ with effective radii of $1$--$20\pc$ and a median of $r_\mathrm{eff}\sim3\pc$, slightly larger in early-type than late-type galaxies \citep[see][for a comprehensive review]{Neumayer2020review}.
Within the Local Group, several NSCs have been studied in considerable detail.
The Milky Way NSC has a mass of $\sim2.5\times10^7\msun$, an effective radius of $\sim4\pc$, and a central density of $\sim2.5\times10^6\msun\pc^{-3}$ within $0.5\pc$ \citep{genzel2010RvMP...82.3121G, 2011ASPC..439..222S, schodel2014A&A...566A..47S, feldmeier2014A&A...570A...2F, 2014CQGra..31x4007S, 2017MNRAS.466.4040F}.
The M31 NSC has a comparable mass of $\sim2.8\times10^7\msun$ enclosed within $\sim12\pc$ \citep{peng2002AJ....124..294P}.
Beyond the Local Group, large samples of NSC masses and sizes have been compiled from Hubble Space Telescope imaging \citep[e.g.,][]{boker2002AJ....123.1389B, georgiev2014MNRAS.441.3570G, 2016MNRAS.457.2122G}.


Spectroscopic studies reveal that NSCs host diverse stellar populations, with luminosity-weighted ages spanning from $\sim10\Myr$ to $\sim10\Gyr$ in both early- and late-type hosts \citep{2006AJ....132.1074R, walcher2006ApJ...649..692W, seth2006AJ....132.2539S}.
Some nearby NSCs in early-type galaxies possess significant populations younger than 100~Myr \citep{monaco2009A&A...502L...9M, kacharov2018MNRAS.480.1973K}, while the Milky Way NSC appears to consist of a dominant old component formed more than $5\Gyr$ ago alongside a younger population with ages $<5\Gyr$ \citep{blum2003ApJ...597..323B, pfuhl2011ApJ...741..108P}.
These age distributions are essential diagnostics for distinguishing among NSC formation scenarios, yet they remain highly degenerate and uncertain due to sparse spectroscopic sampling, the age--metallicity degeneracy, host-galaxy light contamination, and dust extinction \citep{fahrion2021A&A...650A.137F, Neumayer2020review}.


The formation and growth of NSCs are thought to proceed through two primary channels.
In the first---in-situ star formation---gas is funneled to the galactic center through processes such as bar-driven infall, dissipative nucleation, galaxy mergers, and magnetorotational instability \citep{1989Natur.338...45S, 1991ApJ...376..214B, 1995ApJ...448...41H, 2007PASA...24...77B, 2010Natur.466.1082M, antonini2015ApJ...812...72A}.
Once at the center, the gas cools, fragments, and forms stars; repeated episodes of star formation build up the NSC over time \citep{loose1982A&A...105..342L, levin2003ApJ...590L..33L, milosavljevi2004ApJ...605L..13M, bartko2009ApJ...697.1741B, antonini2015ApJ...812...72A, feldmeier-krause2015A&A...584A...2F, nogueras-lara2020NatAs...4..377N}.
This channel naturally explains the young stellar populations frequently observed in NSCs, as well as their complex star formation histories and metallicity spreads \citep{kacharov2018MNRAS.480.1973K, fahrion2019A&A...628A..92F, fahrion2021A&A...650A.137F, pinna2021ApJ...921....8P, hannah2021AJ....162..281H}.
Observational support comes from the detection of young, massive stars near galactic centers that are dynamically distinct from the older NSC population \citep{walcher2006ApJ...649..692W, feldmeier-krause2015A&A...584A...2F, alfaro-cuello2019ApJ...886...57A, fahrion2019A&A...628A..92F}.

The balance between gas accretion and expulsion plays a central role in regulating in-situ star formation within NSCs.
Gas accretion onto NSCs is driven by gravitational interactions, galaxy mergers, or secular processes such as bar instabilities \citep{leigh2014MNRAS.441..919L, karam2024ApJ...967...86K}.
However, the resulting star formation also triggers feedback---stellar winds, supernova explosions, and radiation pressure---that can expel gas and regulate further growth \citep{pelupessy2012MNRAS.420.1503P}.
Despite considerable observational evidence, simulation studies that self-consistently model in-situ star formation within a live NSC embedded in a galactic environment remain scarce.


In the second channel, star clusters that form in the outer regions of galaxies spiral inward under dynamical friction and eventually merge with the NSC \citep{tremaine1975ApJ...196..407T}.
\citet{capuzzo1993ApJ...415..616C} showed that the competition between dynamical friction and tidal disruption shapes the globular cluster population in triaxial elliptical galaxies, with dynamical friction prevailing unless the nuclear mass is sufficiently large.
The observed deficit of massive star clusters in the inner regions of many galaxies supports this migration picture \citep{lotz2001ApJ...552..572L, capuzzo2009MSAIS..13...35C, sanchez2019ApJ...878...18S}, and the presence of metal-poor stars in the nuclei of low-mass galaxies---in contrast to their more metal-rich surroundings---further corroborates globular cluster infall rather than in-situ formation as the origin \citep{paudel2011MNRAS.413.1764P, mucciarelli2017A&A...605A..46M, kacharov2018MNRAS.480.1973K}.

Direct observational evidence for ongoing mergers comes from the detection of NSCs closely surrounded by star clusters.
\citet{georgiev2014MNRAS.441.3570G} identified two adjacent massive star clusters at the center of NGC~4654, and \citet{schiavi2021MNRAS.503..594S} showed via direct N-body simulations that they are fated to merge within $\sim20$--$80\Myr$.
More recently, JWST observations revealed an NSC of mass $2.5\times10^7\msun$ at the center of NGC~4654 flanked by two young star clusters at projected distances of 21.6~pc and 14.3~pc \citep{fahrion2024A&A...687A..83F}.
Similarly, \citet{smith2020ApJ...896...84S} identified three adjacent star clusters in the core of NGC~5253.
Although the cluster infall scenario is well supported, it alone cannot account for the young stellar populations present in many NSCs, since globular clusters are typically older than $\sim1\Gyr$.
Growing evidence suggests that both channels operate simultaneously, with their relative importance varying with galaxy mass and morphological type \citep{fahrion2021A&A...650A.137F, fahrion2022A&A...667A.101F}.


The rotational properties of NSCs provide additional diagnostics for disentangling these formation channels.
Observed NSCs span a wide range of kinematics: some exhibit minimal rotation, consistent with dry accretion of globular clusters on randomly oriented orbits \citep{nguyen2018ApJ...858..118N}, while others show significant angular momentum, which can be attributed to in-situ star formation from a coherently spinning gas reservoir or to coplanar cluster mergers \citep{fahrion2019A&A...628A..92F}.
\citet{2019A&A...629A..44L} surveyed the stellar kinematics of early-type galaxies and their NSCs, finding diverse rotational behaviors and concluding that the pure globular-cluster merging scenario remains viable.
\citet{pinna2021ApJ...921....8P} demonstrated that NSCs in late-type hosts tend to be rotation-dominated, whereas those in early-type galaxies rotate more slowly with lower ellipticity.
However, the complex morphologies and diverse evolutionary histories of host galaxies mean that kinematics alone cannot fully reconstruct NSC assembly; spatially resolved stellar-population studies and detailed star formation histories are essential complements \citep{Neumayer2020review}.


Numerical simulations have been instrumental in testing these formation scenarios, advancing progressively in physical fidelity.
Early efforts focused on the cluster infall channel using pure N-body methods.
\citet{antonini2012ApJ...750..111A} performed direct N-body simulations of 12 globular clusters spiraling into the Milky Way NSC, using $\sim5{,}700$ particles of $200\msun$ per cluster and $240{,}000$ particles of $\sim380\msun$ for the galaxy, with a gravitational softening of 0.01~pc.
\citet{perets2014ApJ...784L..44P} explored the structural signatures of the cluster infall scenario on NSCs and their multiple stellar populations using an analogous setup.
\citet{panamarev2019MNRAS.484.3279P} raised the resolution to $\sim10^6$ particles with 5\% primordial binaries, integrating single and binary stellar evolution to study the Milky Way NSC.
\citet{mastrobuono2021MNRAS.505.3314M} assessed the impact of stellar close encounters on NSC populations, within the merger framework of \citet{antonini2012ApJ...750..111A}.
\citet{schiavi2021MNRAS.503..594S} investigated the evolutionary trajectories of two NSCs in NGC~4654 using direct N-body simulations.
While these studies have provided fundamental insights into the merger channel, they lack hydrodynamics and therefore cannot model in-situ star formation.

More recently, simulations have begun to incorporate hydrodynamics.
\citet{guillard2016MNRAS.461.3620G} tested a wet migration scenario---NSC formation via mergers of gas-rich globular clusters---with hydrodynamic simulations of an isolated dwarf galaxy at 3.5~pc resolution and $700\msun$ star particles.
\citet{gray2024arXiv240519286G} used the EDGE suite of high-resolution dwarf galaxy simulations ($\sim3\pc$, $300\msun$) to discover a new formation mechanism: massive NSCs ($8\times10^5\msun$) forming in low-mass halos ($M_{200}\sim5\times10^9\msun$) at $z\sim2$ through merger-driven starbursts.
\citet{vandonkelaar2024MNRAS.529.4104V} studied NSC formation within 1.5~kpc of Milky Way-mass galaxies at $z>4$ using a cosmological zoom-in simulation at $788\msun$ mass resolution.
\citet{partmann2024arXiv240918096P} investigated the co-evolution of NSCs and massive black holes (MBHs) in dwarf galaxies using hydrodynamic simulations with the KETJU code at $0.5$--$4\msun$ resolution, enhancing the gravitational dynamics around MBHs but without a direct N-body solver.
\citet{lahen2024arXiv241001891L} ran star-by-star hydrodynamic simulations of star cluster formation in low-metallicity dwarf galaxies, enhancing collisional dynamics through adaptive softening and regularization, though without a complete direct N-body solver.
These hydrodynamic studies have greatly expanded our understanding of the galactic environment in which NSCs form, but they generally cannot resolve the internal collisional dynamics of star clusters---precisely the regime where mergers, mass segregation, and core collapse shape NSC structure.

\citet{2024ApJ...974..193J} introduced \newenzo{}, a hybrid code that couples an adaptive mesh refinement hydrodynamic solver with a direct N-body integrator, for the first time resolving galaxies and star clusters concurrently while maintaining collisional accuracy within star clusters.
However, a critical gap remains: no simulation to date has self-consistently modeled both the in-situ star formation channel and the cluster merger channel within a live galaxy that includes collisional stellar dynamics, hydrodynamics, star formation, and supernova feedback simultaneously.


In this work, we develop and integrate a new in-house direct N-body code, \nbody{} (Jo et al., in prep.), into \oldenzo{} \citep{2024ApJ...974..193J}, replacing the previously employed {\tt Nbody6++GPU}. 
We refer to this updated version as \newenzo{}.
\nbody{} is designed to handle complex and adaptive stellar systems within galaxies while simultaneously supporting subgrid physics---such as star formation and supernova feedback---that interact with the hydrodynamics.
Close encounters and few-body dynamics are treated with a Slow-Down Algorithmic Regularization (SDAR) scheme \citep{sdar2020MNRAS.493.3398W}.
This framework enables us to simulate both the host galaxy and the NSC concurrently, incorporating live dark matter, gas, and stars, and to resolve the internal dynamics of star clusters at a level comparable to that achieved by the dedicated direct N-body community.

Using \newenzo{}, we simulate a dwarf galaxy ($M_\mathrm{tot}\sim10^{10}\msun$) hosting a $3\times10^5\msun$ NSC at its center, focusing on how the NSC evolves through both mergers with nearby star clusters and in-situ star formation.
Gas fragmentation in the dwarf galaxy drives the formation of star clusters, some of which migrate inward via dynamical friction and merge with the NSC; concurrently, gas accretion onto the NSC fuels in-situ star formation.
To disentangle the roles of these two growth channels and the influence of stellar physics, we perform four simulations: a fiducial run, one without supernova feedback, one with a lower star formation threshold, and one with twice the galactic gas content.

The remainder of this paper is organized as follows.
Section~\ref{sec:method} describes the \newenzo{} framework, including the hydrodynamic solver (Sec.~\ref{sec:method_hydro}), the stellar physics (Sec.~\ref{sec:method_star_physics}), and the \nbody{} direct N-body solver (Sec.~\ref{sec:method_nbody}).
Section~\ref{sec:method_hdbscan} introduces the \texttt{HDBSCAN} cluster identification method.
Section~\ref{sec:result} presents our results: an overview of the host galaxy and NSC (Sec.~\ref{sec:result_overview}), the evolution of the NSC through mergers (Sec.~\ref{sec:result_merger}), through in-situ star formation (Sec.~\ref{sec:result_insitu}), the gas properties within the NSC (Sec.~\ref{sec:result_nsc_gas}), and the rotational dynamics (Sec.~\ref{sec:result_rotation}).
In Section~\ref{sec:discussion}, we discuss the interplay between mergers and in-situ star formation (Sec.~\ref{sec:discussion_interplay}), numerical validation including core collapse and dark matter resolution effects (Sec.~\ref{sec:discussion_numerical}), and limitations of our approach (Sec.~\ref{sec:discussion_limitations}).
We summarize our findings in Section~\ref{sec:summary} and outline future directions in Section~\ref{sec:future_work}.
In Appendix \ref{sec:appendix_numerical}, we summarize numerical validation tests in full details.

\section{Methodology}
\label{sec:method}

We use {\newenzo}, a cutting-edge hybrid code that integrates a direct N-body simulation into a cosmological (magneto-)hydrodynamic code \citep{2024ApJ...974..193J}.
{\newenzo} comprises two main components: the `hydro' part and the `nbody' part. 
The hydro part, based on Enzo, handles cosmology, (magneto-)hydrodynamics, collisionless gravity, and sub-resolution physics such as radiative cooling, star formation, and black hole accretion. 
On the other hand, the `nbody' part is dedicated to modeling collisional gravity through the direct sum method. 
The interaction between these two components is facilitated by the use of the ``background acceleration'', which is calculated within the `hydro' part using the PM solver and incorporates contributions from dark matter, gas, and stars within the galaxy, excluding those in star clusters.

In this work, we develop and integrate a new direct N-body code \nbody{} (Jo et al. in prep) into \newenzo{} as a replacement of the pre-existing direct N-body solver, {\tt Nbody6++GPU} that was adopted formerly in \citet{2024ApJ...974..193J}.
\nbody{} is sophisticatedly architected to establish a connection with the `hydro' part in both physical and technical dimensions.
With this, we can fully utilize the subgrid models in \enzo{} such as star formation, supernova feedback, black hole accretion and feedback within the `nbody' part.
In other words, star clusters can evolve with a full physics together with their host galaxies.

\subsection{Hydrodynamics}
\label{sec:method_hydro}
{\enzo}---the foundation for \newenzo{}---is a multi-purpose gravito-hydrodynamic computational code designed to tackle a broad range of astrophysical problems from large-scale structure formation to the evolution of black holes \citep{enzo2014ApJS..211...19B}. 
The framework possesses a comprehensive array of subgrid physics, including star formation and feedback, radiative cooling, and black hole accretion.
{\enzo} employs an Eulerian scheme built upon block-structured adaptive mesh refinement as its hydrodynamic engine, encompassing four distinct numerical hydrodynamic techniques: the hydrodynamic-only piecewise parabolic method \citep{colella1984JCoPh..54..174C,bryan1995CoPhC..89..149B}, the MUSCL-like Godunov scheme \citep{vanleer1977JCoPh..23..276V,wang2008ApJS..176..467W,wang2009ApJ...696...96W}, a constrained transport (CT) staggered magnetohydrodynamics (MHD) scheme \citep{collins2010ApJS..186..308C}, and a second-order finite difference hydrodynamics method \citep{stone1992ApJS...80..753S,stone1992ApJS...80..791S}.
Simultaneously, gravitational forces are addressed through the application of the Particle Mesh solver (PM solver), in conjunction with the mesh structure.
The PM solver advances collisionless particles through a singular timestep via a drift-kick-drift algorithm \citep{hockney1988csup.book.....H}. 
This methodological approach guarantees second-order accuracy, accommodating varying timesteps induced by the dynamic grid structures inherent in adaptive refinement.

\subsubsection{Time-stepping in Hydrodynamics}
The criteria for time-stepping include considerations of hydrodynamics, acceleration, particle velocity, radiation pressure, heat conduction, and the cosmological expansion of the universe, as detailed in Section 9 of \citet{enzo2014ApJS..211...19B}.
The hydrodynamic time step is given as
\begin{equation}
    \Delta t_{\rm hydro} = \min\left(\kappa_{\rm hydro}\left(\sum_{x,y,z}\frac{c_{\rm s}+|v_{x}|}{a\Delta x}\right)^{-1}\right)_{l},
\end{equation}
where $c_s$, $a$, $\Delta x$, and $\kappa$ correspond to the sound speed, the cosmic scale factor, the cell size, and a dimensionless constant that ensures compliance with the Courant–Freidrichs–Levy condition, respectively. 
$\min(\cdots)_l$ determines the minimum value for all cells or particles at a specific level $l$. 
Here, `level' refers to resolution of a simulation. 
The size of a cell (i.e., spatial resolution) at a level $l$ is given by $\Delta x=(W/2)^l$ where $W$ is the size of the root grid of simulations (refer to Sec.~\ref{sec:method_resolution}).
In this study, we utilize $\kappa_{\rm hydro}$ with a value of 0.3.
For gravitational acceleration, the time step is determined by 
\begin{equation}
    \Delta t_{\rm acc} = \min\left(\sqrt{\frac{\Delta x}{|\boldsymbol{g}|}}\right)_{l}.
\end{equation}
The integration of the time step with the drift-kick-drift algorithm ensures second-order precision in gravity \citep{hockney1988csup.book.....H}.

\subsection{In-House Direct N-body: {\nbody}}
\label{sec:method_nbody}
We implement a novel in-house direct N-body code, \nbody{}, in substitution for {\tt Nbody6++GPU}, to attain flexibility and seamless integration with \enzo{}.
\nbody{} adopts the block time step and the fourth order Hermite integrator, along with the Ahmad-Cohen neighbor scheme (see Sec. \ref{sec:method_ac-scheme}).
The few-body dynamics is solved with the modified Slow-Down Algorithmic Regularization (SDAR) \citep{sdar2020MNRAS.493.3398W}.
The details will be described in Jo et al. in prep.

\subsubsection{Hermite integration with block time steps}
\label{sec:method_hermite}
We adopt the fourth-order Hermite integration scheme described in \citet{makino1991ApJ...369..200M} and \citet{aarseth2003gnbs.book.....A}.
This method utilizes the lower order acceleration derivatives---$\bs{a}_0, \bs{a}, \bs{a}_0^{(1)}, \bs{a}^{(1)}$---to obtain approximate values for higher order acceleration derivatives---$\bs{a}^{(2)}, \bs{a}^{(3)}$---of each particle (see Eq. 2.20 and 2.21 in \citet{aarseth2003gnbs.book.....A}). 
A Taylor series expansion of the acceleration and its first derivative is expressed to third order with respect to the reference time $t$ as follows:
\begin{equation}
\begin{split}
   \bs{a}&=\bs{a}_0+\bs{a}^{(1)}_0t+\frac{1}{2}\bs{a}^{(2)}_0t^2+\frac{1}{6}\bs{a}^{(2)}_0t^3,\\
   \bs{a}^{(1)}&=\bs{a}^{(1)}_0+\bs{a}^{(2)}_0t+\frac{1}{2}\bs{a}^{(3)}_0t^2.
\end{split}
\end{equation}
Substituting $\bs{a}^{(2)}_{0}$ and $\bs{a}^{(3)}_{0}$, we obtain 
\begin{equation}
\begin{split}
   \bs{a}^{(2)}_{0}&=\frac{6}{t^3}\left[2\left(\bs{a}_0-\bs{a}\right)+\left(\bs{a}^{(1)}_0+\bs{a}^{(1)}\right)t\right],\\
   \bs{a}^{(3)}_{0}&=\frac{2}{t^2}\left[-3\left(\bs{a}_0-\bs{a}\right)-\left(2\bs{a}^{(1)}_0+\bs{a}^{(1)}\right)t\right].
\end{split}
\end{equation}
where $\mathbf{a}^{(n)}$ is the n-th derivative of the acceleration.

Block time steps are allocated to individual particles such that the quantized time step for each particle is smaller than the time step criteria. 
The criteria of time steps for the particles is given by
\begin{equation}
\label{eq:tcrit}
    \Delta t_\mathrm{Nbody} = \sqrt{\eta \frac{ |\mathbf{a}^{(0)}| |\mathbf{a}^{(2)}| + |\mathbf{a}^{(1)}|^{2} }{ |\mathbf{a}^{(1)}| |\mathbf{a}^{(3)}| + |\mathbf{a}^{(2)}|^{2} } },
\end{equation}
where $\Delta t$ is the criteria for time step \citep{aarseth2003gnbs.book.....A} and the steps are quantized in powers of 1/2. 
The time-stepping is independently managed within the 'hydro' and 'nbody' parts such that the time steps of the `nbody' part retain their consistency regardless of communication with the `hydro' part, ensuring that the physics remains unaffected (refer to Sec. 2.2.1 of \citet{2024ApJ...974..193J} for details).

\subsubsection{Ahmad-Cohen neighbor scheme}
\label{sec:method_ac-scheme}

Direct calculation of particles up to fourth-order is computationally expensive, so {\nbody} employs a neighbor scheme described by \citet{ahmad1973JCoPh..12..389A}. 
In this scheme, the total force on a particle is divided into short-range ({\it irregular}) and long-range ({\it regular}) components. 
The irregular force is calculated from neighbor particles within a certain radius (the AC radius) of the target particle. 
All the other particles are considered as contributors to the regular force. 
The irregular force is updated at shorter intervals than the regular force. 
During irregular time steps, the regular force is estimated up to first-order using the derivatives calculated at regular time steps. 

Within this framework, the choice of the radius which encompasses the neighbors influences both computational cost and the accuracy of simulations. 
However, the optimization of this remains unresolved. 
In this work, we adopt two distinct approaches.
Initially, during the simulations, neighbors are identified employing the N-nearest neighbors method to determine the AC radii of each particle that includes N-nearest neighbors. 
Subsequently, the AC radius is updated based on the number of neighbors by implementing penalization or reward based on this number. 
This approach ensures the stabilization of the neighbor particle count for each particle.

\subsubsection{Close encounters}
\label{sec:method_regularization}

Accurate modeling of close encounters is crucial for the correct evolution of star clusters, especially after core collapse. 
The Hermite scheme for managing close encounters is inefficient in computational time and lacks precision. 
Therefore, \nbody{} is implemented with the modified SDAR package to handle close encounters.
However, in this study, we have opted not to consider the physics of close encounters due to the limitations imposed by particle mass resolution. 
Each star particle in our simulations has a mass of about $300\,M_\odot$, with each particle standing in for a stellar population (refer to Sec.~\ref{sec:method_star_physics}).
Given these conditions, if a star particle were to participate in close encounter interactions, such as two-body or three-body encounters, the resultant `group' of stars might either result in the formation of a binary system or be expelled from the star cluster, which is an undesirable outcome.
The impact of this choice will be discussed in Sec.~\ref{sec:appendix_core_collapse}.

\subsubsection{Direct N-body Region}
\label{sec:method_direct_nbody_region}

To balance computational cost against dynamical fidelity, we restrict the collisional (direct) N-body solver to a time-dependent spherical domain of radius $R_\mathrm{N\,body}=200\pc$, centered on the center of mass of the NSC at each time step.
Particles that exit this domain are transferred back to the collisionless particle-mesh (PM) solver, while particles that enter it are promoted to the direct integrator.
This hybrid scheme avoids the $\mathcal{O}(N^2)$ cost of a full-volume direct integration while preserving collisional accuracy where it matters most---in the immediate vicinity of the NSC.

The choice of $R_\mathrm{N\,body}=200\pc$ is motivated by the dynamical friction timescale.
For star clusters of mass $\sim10^5\msun$ orbiting within this radius, the inspiral timescale is $\sim20$--$40\Myr$ (see Sec.~\ref{sec:result_merger}), ensuring that all clusters relevant to NSC growth through mergers are captured by the direct solver within the 150~Myr simulation runtime.
Clusters forming at larger radii ($r>1\kpc$) would require $\sim\mathrm{Gyr}$ to reach the nucleus and do not contribute to NSC evolution on the timescales accessible to our simulations.
Our tests confirm that the $200\pc$ aperture captures virtually all globular clusters that sink to the nucleus and merge with the NSC within the simulation time.

It is important to note that the resolution of the direct N-body solver is completely independent of the hydrodynamic grid resolution: the gravitational interactions between star particles within $R_\mathrm{N\,body}$ are computed via direct summation with no spatial discretization, while the hydrodynamic quantities are evolved on the adaptive mesh refinement grid \citep[for details, see][]{2024ApJ...974..193J}.
This decoupling ensures that the internal dynamics of star clusters---including core collapse, mass segregation, and tidal interactions during mergers---are resolved at collisional accuracy regardless of the gas cell size.

\subsection{Resolution of Simulations}
\label{sec:method_resolution}
We refine the simulation box of $(1.31\,{\rm Mpc})^3$ both {\it dynamically} and {\it statically} up to level 15, which corresponds to the finest cell size of $0.625\,{\rm pc}$. 
A {\it static} refinement region is employed to achieve a resolution of the $(5 \kpc)^3$ box to 160 pc, which encompasses the entire host galaxy.
For the further refinement, the cells are {\it dynamically} refined, following two different refinement criteria: 1) gas mass of a cell; 2) mass of the particles contained in a cell.
Notice that these criteria are applied in an “OR” fashion; a cell is refined whenever it satisfies at least one of the two conditions.
The mass thresholds for gas and particles above which a cell is refine at level $l$ is as follows:
\begin{align}
     M^{l}_{\rm ref, gas} &= M_{\rm ref, gas, min}\, r^{\epsilon l}\\
     M^{l}_{\rm ref, part} &= M_{\rm ref, part, min}\, r^{\epsilon l} 
\end{align}
 where $M_{\rm ref,min}$, $\epsilon$, and $r$ are the minimum mass for refinement, a refinement exponent, and a refinement factor which is 2 conventionally. 
`part' and `gas' denote the mass of particles and gas cells, respectively.
 Here, $M_{\rm ref, min}$ and $\epsilon$ are the free parameters that control the degree of refinement.

In the case of gas, to be consistent with the star formation scheme (Section \ref{sec:method_star_physics}), we set the mass threshold for gas at the finest level $l_{\rm max}$ to be the order of the Jeans mass $M_{\rm J}$ that relates to fragmentation due to the gravitational instability: i.e., $M^{l_{\rm max}}_{\rm ref,gas} \sim M_{\rm J}$ \citep{truelove1997ApJ...489L.179T}. 
The Jeans mass of the finest cell at $l=15$ is approximately $300 \msun$ assuming the Jeans length is the size of a cell $\Delta x_{15} = 0.625 \pc$ with $T=100K$. 
We choose $M^{15}_{\rm ref, gas} = 300 \msun$ and $\epsilon = -0.5$ so that the cells can be refined more on smaller scales (super-Lagrangian).
In terms of particles, the mass criterion requires the refinement of a cell when it encompasses a particle mass that is approximately equivalent to four star particles.

\begin{table}
    \centering
    \begin{tabular}{c|c|c|c}
        Simulation &  Feedback\footnote{Supernova feedback} & $n_\mathrm{thres}$\footnote{Number density threshold for star formation in $\text{cm}^{-3}$ (see Sec.~\ref{sec:method_star_physics})} & $M_\mathrm{gas}$\footnote{Mass of a galaxy gaseous disc in $\msun$ (see Sec.~\ref{sec:method_IC})} \\
        \hline
        \hline
        Fiducial &  on &{77000}& $1.13\times10^8$\\
        \hline
        No Feedback & {\bf off} &{77000}& $1.13\times10^8$\\
        \hline
        Low Star Formation & on &{\bf 38500}& $1.13\times10^8$\\
        \hline
        $2\times$ Gas & on  &{77000}& $\boldsymbol{2.26\times10^8}$
    \end{tabular}
\caption{List of the simulations and the descriptions.}
\label{tab:sim_params}
\end{table}

\subsection{Star Formation and Thermal Feedback}
\label{sec:method_star_physics}
We adopt the star formation recipe that is built based on the work of \citealp{cen&ostriker1992ApJ...399L.113C} with several modifications from \citealp{kim2011ApJ...738...54K, kimI2013ApJ...775..109K, kimII2013ApJ...779....8K}.  
The star particle is created in the center of the cell whenever the cell satisfies the following criteria: 
1) a cell is refined to the finest level; 
2) the proton number density exceeds the threshold $n_{\rm thres} = 77000 {\rm \,cm^{-3}}$;
3) the velocity is converging $\nabla \cdot v < 0$;
4) the gas in the cell cools rapidly enough: $t_{\rm cool} < t_{\rm ff}$ and $T < 1.1 \times 10^4 {\rm K}$.
$n_{\rm thres}$ is estimated by the Jeans mass as in the previous section.
Once the criteria are satisfied, the cell produces a star particle with mass $M^{\rm init}_{} = \epsilon_{\star}\rho_{\rm cell}\Delta x_{15}^{3}$ where the star formation efficiency $\epsilon_{\star}$ is set to 0.5 for the fiducial simulation.

The star particle evolves as follows \citep{butsky2017ApJ...843..113B}:
 \begin{equation}
 \label{eq:star_evolution}
 \begin{split}
     M_{\star}(t) &= \eta_\mathrm{SN} M_\mathrm{init}\int_0^{\tau} \tau'e^{-\tau'}d\tau'\\
     &=\eta_{SN} M_\mathrm{init}[1-(1+x\tau)e^{-\tau}],
 \end{split}
\end{equation}
where $\tau \equiv (t - t_{\rm cr})/t_\mathrm{dyn}$ with a creation time $t_\mathrm{cr}$ and a dynamical time $t_\mathrm{dyn}$, and $\eta_\mathrm{SN}$ determines the fraction of mass that should be locked up after the supernova feedback, which is set to 0.75 in this work.
The supernova feedback occurs during 12 $t_\mathrm{dyn}$. 
At its birth, $0.25 M_{\rm init}$ is subject to supernova feedback that is returned back to the gas cells. 
This process converts the rest mass of the ejected gas into thermal energy ($E=mc^2$) with a factor of $7.5\times10^{-8}$, equivalent to $\sim10^{49} \erg$ per $75\msun$, which is $25\%$ of $300\msun$.
The selected conversion factor yields a lower energy budget for supernovae than the typical value of $\sim 10^{51}\erg$ per $100\msun$.
Nevertheless, the conversion factor is selected based on empirical observations such that the integrity of the host galaxy is maintained.

The evolution of mass within the 'nbody' part is performed independently, due to the discrepancy in time steps between the hydrodynamics and the direct N-body solver. 
Specifically, the stars within the 'nbody' part evolve according to the identical equations, Eq. \ref{eq:star_evolution}, with the 'nbody' time steps and their creation times. 
Subsequently, during each instance where the synchronization of the 'hydro' and 'nbody' segments is achieved, mass synchronization is correspondingly effected.
Consequently, the mass evolution of stars has an impact on the gravitational dynamics of stars in real time. 

\begin{table}[t!]
\centering
\begin{tabular}{lr}
  \hline
  \hline
  \parbox{4.3cm}{\bf Dark matter halo}  &\parbox{3.5cm}{} \\
  \hline
  Virial mass  & $9.56\times10^9\msun$ \\
  Particle mass ($\msun$) & $10^4$ \\
  Profile & Navarro-Frenk-White \\
  Concentration & 13.2\\
  Scale radius (kpc) & 3.40\\
  \hline
  \textbf{Disk star} \\
  \hline
  Total mass & $9.03\times10^8$ \\
  Particle mass ($\msun$) & $10^3$ \\
  Profile & Exponential \\
  Scale radius (kpc) & 1.48 \\
  Scale height (kpc) & 0.148 \\
  \hline
    \textbf{Bulge star} \\
  \hline
  Total mass & $4.52\times10^8$ \\
  Profile & Hernquist \\
  Particle mass ($\msun$) & $10^3$ \\
  Scale radius (kpc) & 3.40\\
  \hline
  \textbf{Disk gas} \\
  \hline
  Total mass & $1.13\times10^8$ \\
  Profile & Exponential \\
  Scale radius (kpc) & 1.48\\
  Scale height (kpc) & 0.148\\
  \hline
  \textbf{NSC star} \\
  \hline
  Total mass ($\msun$) & $3\times 10^5$  \\
  Particle mass ($\msun$) & 300 \\
  Profile & Plummer \\
  Effective radius (pc) & 2\\
 \hline
\end{tabular}
\label{tab:galaxy_IC}
\caption{The initial conditions for the host dwarf galaxy and the NSC}
\end{table}

\subsection{Initial Conditions for Dwarf Galaxy and Nuclear Star Cluster}
\label{sec:method_IC}
We generate the initial condition of the host dwarf galaxy with the NSC using the Disk Initial Conditions Environment (DICE) package \citep{dice2016ascl.soft07002P}.
The dwarf galaxy is composed of a live dark matter(DM) halo, stellar disk and bulge, exponential gaseous disk, and the NSC at the center. 
Tab.~\ref{tab:galaxy_IC} summarizes the details of each component.
Dark matter particles in a halo are sampled from an analytic form that follows the Navarro-Frenk-White profile \citep{navarro1997ApJ...490..493N} with a concentration parameter $c = 13.2$, $M_\mathrm{vir} = 9.56\times 10^{9}\msun$, spin parameter $\lambda = 0.04$, and scale radius $R_{\rm s} = 3.4 \kpc$.
The galaxy includes an exponential disk with $M_{\rm disk} = 9.03 \times 10^{8} \msun$, scale length $r_{\rm s} = 1.48 \kpc$ and scale height $h_{\rm s} = 0.1r_{\rm s}$ that comprises 90\% stars and 10\% gas.
Also, a stellar bulge with $M_{\rm bulge} = 4.52 \times 10^8 \msun$ that follows the Hernquist profile is situated. 
In this work, all star particles---including those in the NSCs---are initialized with an age of zero at the start of the simulations, yet they do not produce stellar feedback.

We have inserted the NSC at the center of the host galaxy, generated using the \textsc{McLuster} package \citep{2011MNRAS.417.2300K}. 
The NSC follows a Plummer profile as a function of radius $r$ given as 
 \begin{equation}
    \rho(r) = \frac{3M}{4\pi a^3}\left(1+\frac{r^2}{a^2}\right)^{-\frac{5}{2}},
 \end{equation}
where $\rho$ represents a volume density, $M$ is the total mass of the cluster, and $a$ is the scale length. 
According to the review paper \citet{Neumayer2020review}, there is no consensus on the empirical scaling relation between the effective radius and mass of NSC when the NSC mass is below $10^6\msun$ (see the Fig. 7 in \citet{Neumayer2020review}).
We use an effective radius of $2\pc$, corresponding to $1.53\pc$ in terms of scale length. This choice falls between the empirical NSC size–mass relations for late- and early type galaxies proposed by \cite{2016MNRAS.457.2122G}.

The origin and formation pathways of NSCs remain elusive and are subjects of ongoing investigation. 
Various mechanisms---such as in-situ star formation and the infall and merging of globular clusters---have been proposed, but no single theory has yet achieved consensus. 
Provided with this uncertainty, this work does not commit to any specific formation scenario for NSCs.
Instead, we assume the prior existence of a NSC located at the center of the host galaxy.
Thus, the initial conditions include a pre-existing NSC, and our analysis proceeds without modeling the physical processes that led to its formation.

Although this work has not targeted a particular observed galaxy, we provide a couple of observational counterparts that share resemblance with our galaxy in terms of macroscopic features.
First, NGC 5253 is a blue compact dwarf galaxy with the dynamical mass of $8.3\times10^9\msun$, the stellar mass of $\sim1.14\times10^9\msun$ \citep{2010A&A...521A..63L}, and the gas mass of $\sim5.9\times10^8\msun$ \citep{2012MNRAS.419.1051L}.
The NSC situated at the center of NGC 5253 appears to be the mass of $\sim 2.5\times 10^5\msun$ and have a companion star cluster of $\sim 7.5\times 10^4\msun$ with the separation of a few parsecs \citep{smith2016ApJ...823...38S,smith2020ApJ...896...84S}.
This object shows very strong resemblance to our galaxy.

Another possible reference point is NGC 205.
NGC 205 is an early-type dwarf galaxy with the dynamical total mass of $2\times10^9\msun$ \citep{geha2006AJ....131..332G}, the estimated gas mass of $1.3\times 10^7-4.8\times10^8\msun$ \citep{2012MNRAS.423.2359D}, the stellar mass of $\sim1.1\times10^8\msun$ \citep{2025arXiv250319843L}, and the NSC of $\sim 1.9\times10^5\msun$ \citep{monaco2009A&A...502L...9M}.
Despite the similarities, we again have not aimed for an apple-to-apple comparison between observations and simulations.
The further physical interpretation and comparison will be discussed throughout the paper.


\subsection{Cluster Identification}
\label{sec:method_hdbscan}
We use the Hierarchical Density-Based Spatial Clustering of Applications with Noise package ({\tt HDBSCAN}) to identify star clusters in the simulations \citep{hdbscanmcinnes2017hdbscan}.
\texttt{HDBSCAN} is a clustering algorithm that extends the capabilities of the traditional DBSCAN algorithm. 
It is designed to identify clusters within data by focusing on regions where data points are densely packed together. 
\texttt{HDBSCAN} constructs a hierarchy of clusters, allowing for a more nuanced analysis of the data at different levels of granularity. 
This hierarchical approach is particularly beneficial when dealing with datasets where clusters have varying densities, as it can more accurately capture the complexity of the data.

However, {\tt HDBSCAN} demonstrates a diminished robustness when identifying merging star clusters that are in close proximity to each other.
When a nearby cluster is present, the density excess detected by \texttt{HDBSCAN} becomes less pronounced, which can lead to fewer stars being identified by the cluster finder, particularly in cases where the cluster is undergoing a merger. 
To address these numerical artifacts in the cluster's mass, we post-process the results. 
Starting with the initial conditions of the inserted cluster, we calculate the maximum distance from the center, denoted as $R_{\rm max}$ of star particles. 
In the subsequent snapshot, we first identify the member star particles of the target cluster using \texttt{HDBSCAN}. 
Additional star particles are then included in the target cluster if they meet all of the following criteria: (i) they are not assigned to any cluster in the current time step, (ii) they were part of the target cluster in the previous snapshot, and (iii) they are located within $R_{\rm max}$ from the cluster’s center. 
We update $R_{\rm max}$ based on the updated cluster and repeat this procedure in each subsequent snapshot. 
This adjustment mitigates the unphysical changes in both the radius and mass of the clusters.
However, the \texttt{HDBSCAN} shows a limitation in determining the exact time of the merger, leading to an earlier merging phase. 
In other words, the actual 'merging process' occurs several million years in the simulations after the satellites have been merged into the host within the \texttt{HDBSCAN} framework.

We use \texttt{HDBSCAN} with parameter choices of \texttt{min\_cluster\_size} = 55 and \texttt{min\_samples} = 40, following \citep{2023MNRAS.524..555A}, which provide good reliability upon visual inspection.

Lastly, throughout this paper, any quantity denoted with `NSC'---such as gas mass and dark matter properties of the NSC---refers explicitly to the quantity defined within the radius of NSC which are determined by the \texttt{HDBSCAN} cluster finder.

\subsection{Terminology and definitions}
\label{sec:method_terminology}
{
\setlength{\parindent}{0pt}
{\it Star particle}: A single point particle representing a stellar population of birth mass $\sim 100 \msun$ (see Sec.~\ref{sec:method_star_physics}). \\
{\it Dark matter}: A single point particle representing a collection of dark matter particles with a fixed mass of $10^4\msun$.
Dark matter is not subject to the direct N-body solver, but only resolved with the collisionless PM gravity solver.\\
{\it NSC properties/quantities}: We define every quantity associated with NSCs---for example, the mass of NSC gas, the dark matter content enclosed, and the NSC star formation rate---inside the spherical volume whose radius $R_{\rm max}$ is identified at each time step by {\tt HDBSCAN} (see Sec.~\ref{sec:method_hdbscan}).
Although an NSC is initially inserted at the galactic center on top of the stellar disc and bulge, the surrounding stellar density is sufficiently low that the inserted NSC stars do not overlap with the pre-existing disc or bulge stars.\\
Lastly, throughout this work, the terms {\it Cluster} and {\it Merger} always denote, respectively, a {\it star cluster} and a {\it star cluster merger}, unless explicitly noted otherwise.}


\section{Results}
\label{sec:result}
We simulate an NSC of $M_\star = 3\times10^5\msun$ at the center of a dwarf galaxy of $M_\star\sim10^9\msun$, \tr{unprecedentedly} including live dark matter, gas, and stars, together with star formation and feedback \tr{on top of collisional stellar dynamics.}
By utilizing the direct N-body solver within \newenzo{}, we resolve the kinematics of the NSC in terms of its structural evolution and mergers with nearby clusters.
However, due to computational constraints, the computational domain for the direct N-body calculations is confined to a spherical region centered on the center of mass of the NSC, with a radius of 200 pc (refer to Sec.~\ref{sec:method_direct_nbody_region} for further discussion).

\tr{We have carried out four simulations with different configurations as listed in Tab.~\ref{tab:sim_params}: (1) a fiducial run, (2) without supernova feedback, (3) with low star formation efficiency, and (4) with twice the gas content in the galaxy disc.
With these simulations, we aim to investigate the interplay between different physical processes---such as star formation, supernova feedback, and star cluster mergers---in the evolution of NSCs; e.g., comparison between the fiducial run and the run without supernova feedback provides insight into the effect of supernova feedback on NSC growth (refer to \ref{sec:discussion_limitations}).
}

We first describe the overall features of the host galaxy and the NSC in Sec.~\ref{sec:result_overview}.
In Sec.~\ref{sec:result_merger}, we investigate the impact of mergers on the evolution of the NSCs, while Sec.~\ref{sec:result_insitu} focuses on the contribution of in-situ star formation. 
The gas properties within the NSC are analyzed in Sec.~\ref{sec:result_nsc_gas}, and the rotational dynamics of the NSC are presented in Sec.~\ref{sec:result_rotation}.

\begin{figure*}
    \centering
    \includegraphics[width=0.99\textwidth]{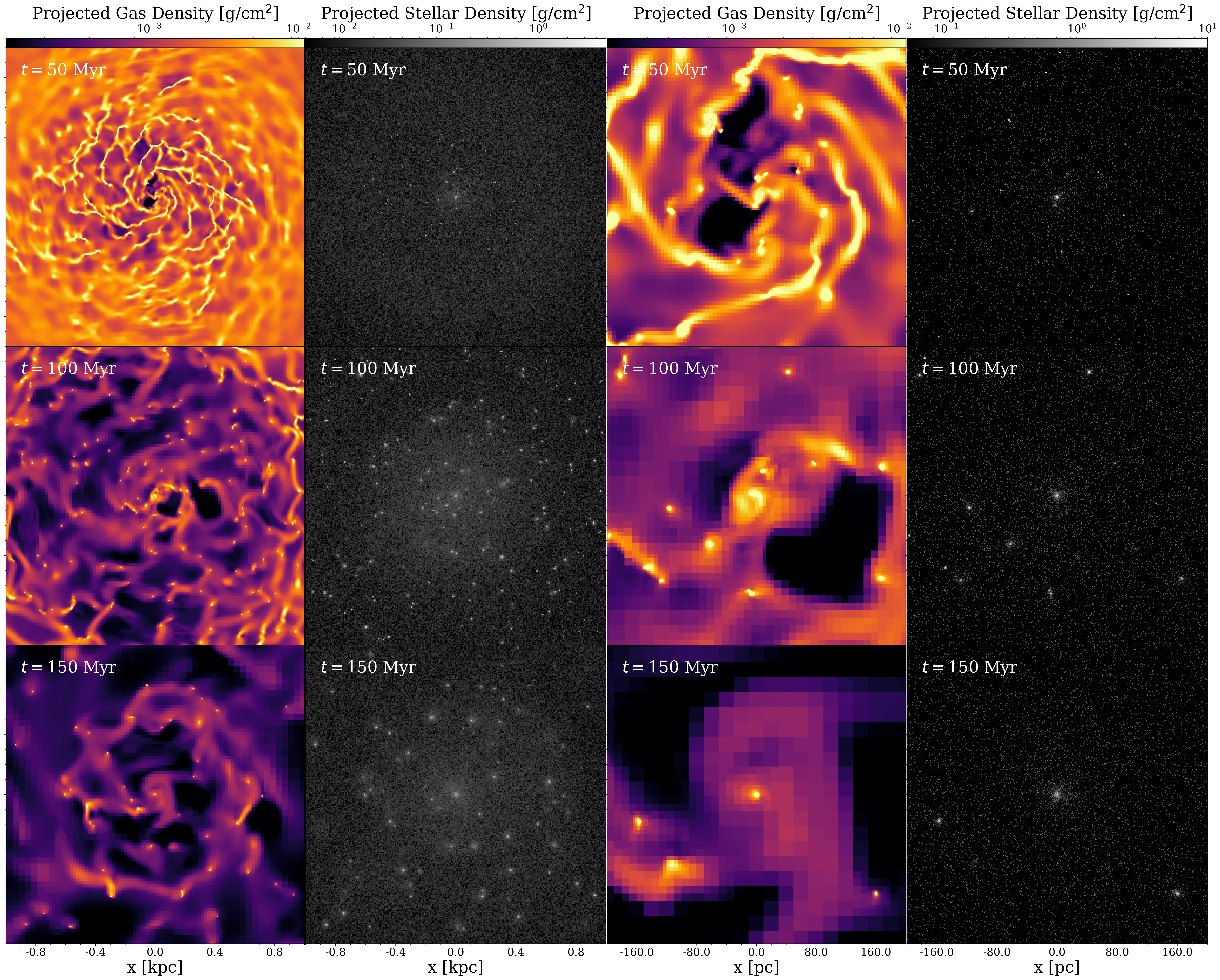}
    \caption{Two dimensional \tr{disk-on} projections of stars and gas in two different views---a galaxy view of a 2 kpc box ({\it first} and {\it second} columns) and a cluster view of a 400 pc box ({\it third} and {\it fourth} columns)---at $t =$ 50 Myr ({\it top}), 100 Myr ({\it middle}), and 150 Myr ({\it bottom}).}
    \label{fig:projection}
\end{figure*}

\begin{figure*}
    \centering
    \includegraphics[width=0.99\linewidth]{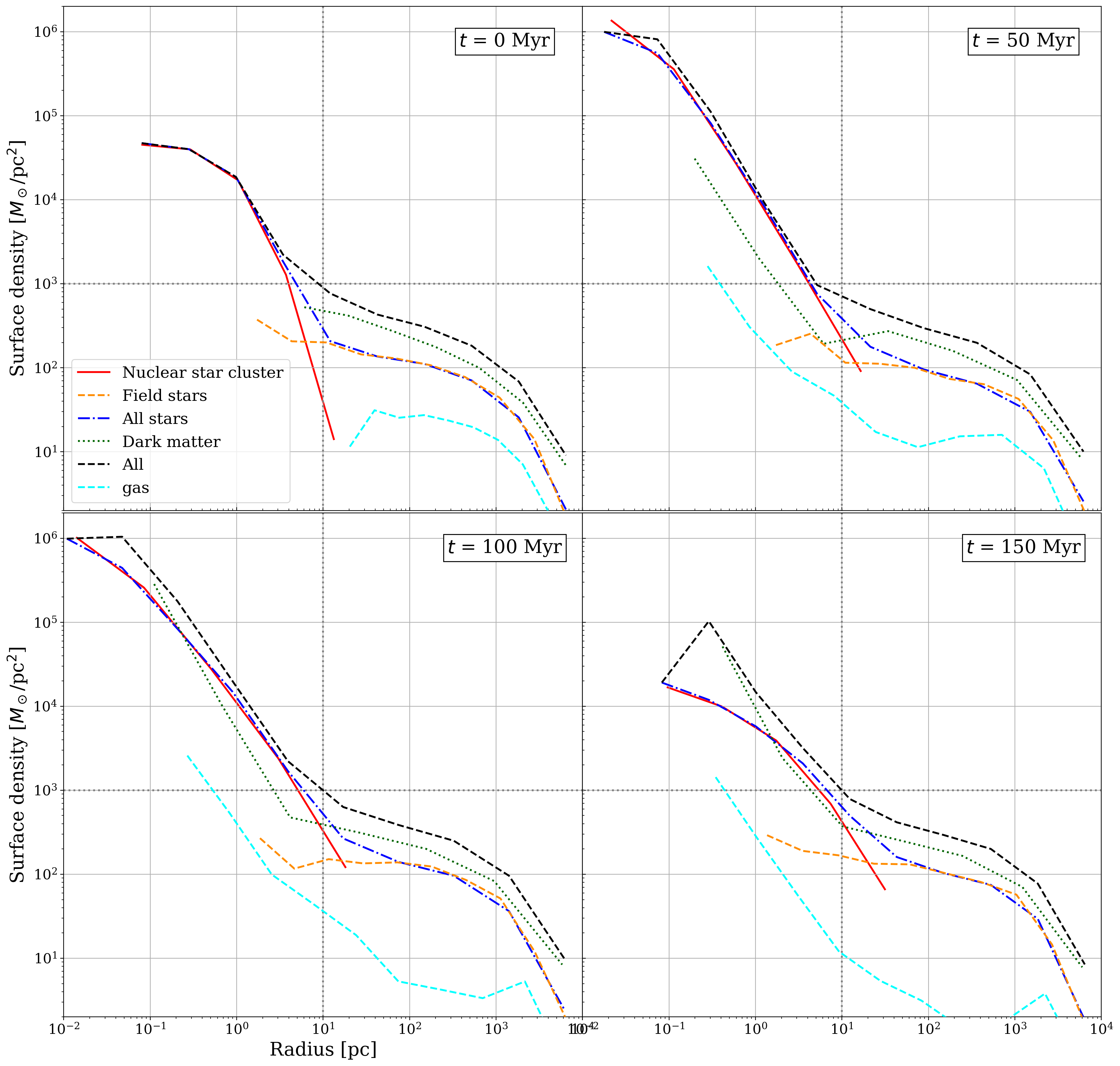}
    \caption{Surface density of the \tr{NSC}, stars and dark matters at 0 Myr ({\it upper left}), 50 Myr ({\it upper right}), 100 Myr ({\it lower left}), and 150 Myr ({\it lower right}).
    The guidelines ({\it grey dotted}) are at 10 pc and $10^3 \msun/\mathrm{pc}^2$.
    After 50 Myr, the \tr{NSC} have become denser and concentrated.
    The dark matters have been able to fall towards the center of the NSC (refer to Sec.~\ref{sec:appendix_dark_matter} for further discussion on the dark matter profiles).
    The recent merger at around 140 Myr results in the disrupted inner profile of NSC at 150 Myr despite the mass growth (refer to Fig. \ref{fig:mass_growth_merger}).
    }
    \label{fig:profile}
\end{figure*}

\begin{figure}
    \centering
    \includegraphics[width=0.99\linewidth]{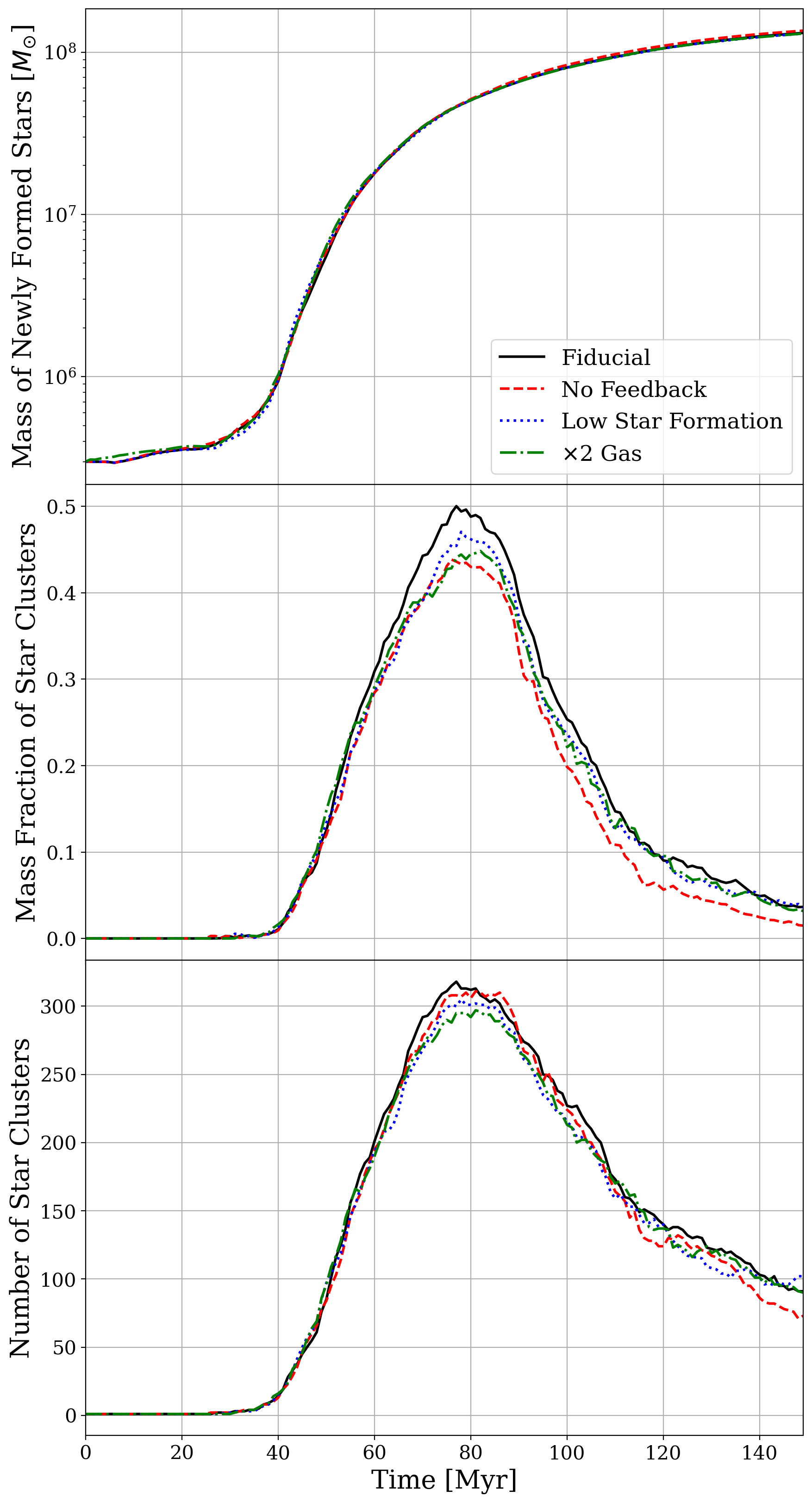}
    \caption{{\it Top}: the temporal evolution of the total mass of newly formed stars within the host galaxy, excluding the initial condition.
    {\it Middle}: the ratio of the aggregate mass of star clusters to the total mass of newly formed stars over time.
    {\it Bottom} the total number of star clusters within the host galaxy as a function of time. 
   The fiducial, no feedback, low star formation, \tr{and double gas mass} simulations are represented by {\it black solid}, {\it red dashed}, {\it blue dotted}, and {\it green dot-dashed} lines, respectively.}
    \label{fig:host_properties}
\end{figure}

\begin{figure}
    \centering
    \includegraphics[width=1.00\linewidth]{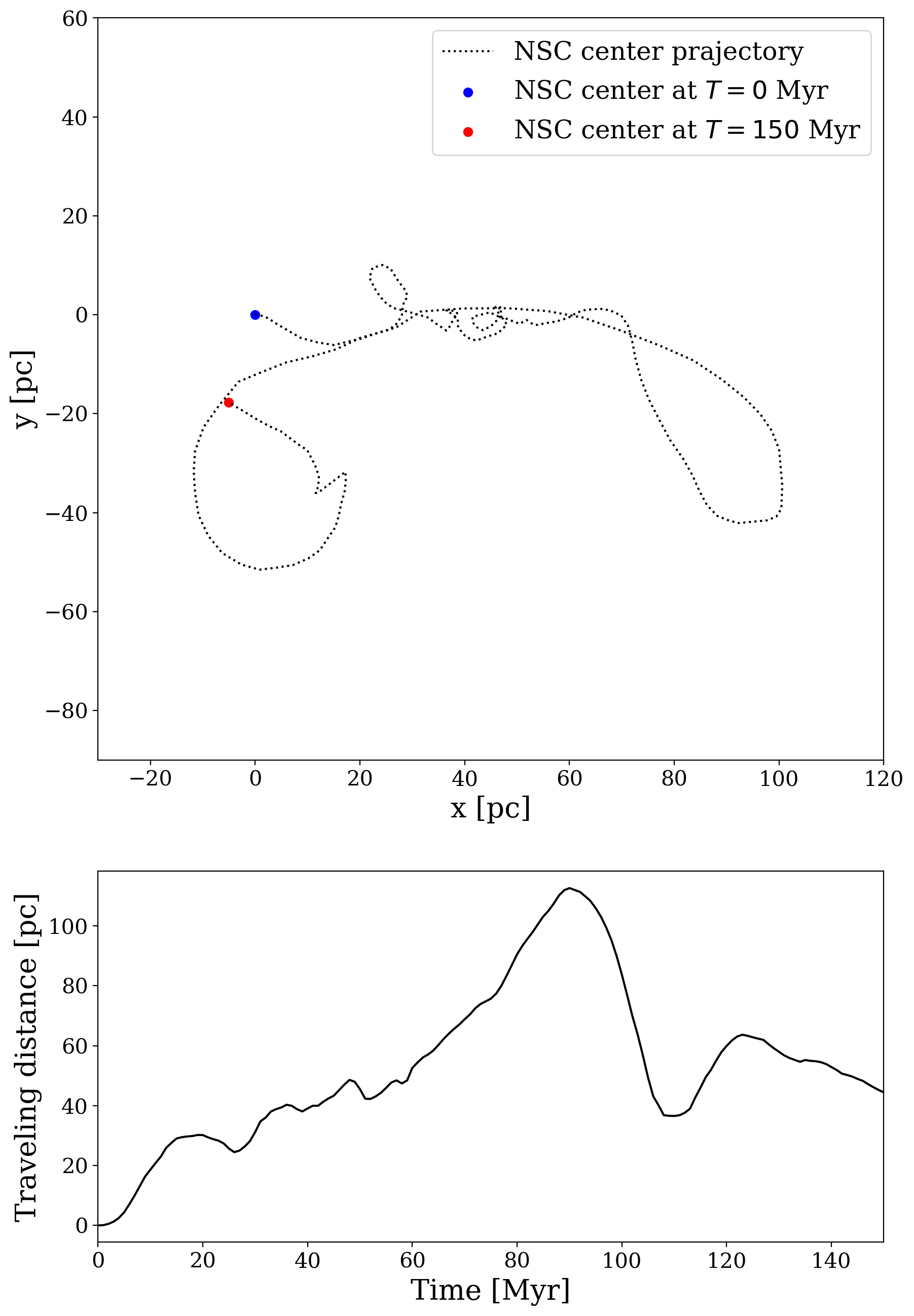}
    \caption{Trajectory of the center of mass of the NSC over 150 Myr ({\it top}). 
    The distance between the centers of mass of the NSC and the host galaxy with time ({\it bottom}).
    The initial position of the center of mass of the NSC at 0 Myr and the final position at 150 Myr are represented with the {\it blue} and {\it red} dot, respectively.
    The two points are connected through trajectory of the center of mass of the NSC ({\it black dotted}).
    The initial position coincides with the center of mass of the host galaxy.
    }
    \label{fig:trajectory}
\end{figure}

\subsection{Overview of Host Dwarf Galaxy and Nuclear Star Cluster}
\label{sec:result_overview}

\subsubsection{Structural Properties and Dynamical Stability}
\label{sec:result_structure}

Fig.~\ref{fig:projection} displays the two-dimensional projections of gas and stars from a galaxy view (2 kpc box) and a star cluster view (400 pc box) at 50 Myr ({\it top}), 100 Myr ({\it middle}), and 150 Myr ({\it bottom}).
The galaxy view ({\it first} and {\it second} columns) illustrates the evolution of the host dwarf galaxy over 150 Myr. 
At the beginning of the \tr{simulations}, \tr{the galaxy initialized with the analytic profiles undergoes a transition phase into more realistic structures through e.g., gas fragmentation.}
Particularly, the gas structures fragment into gas clumps between 50 Myr and 100 Myr, resulting in the formation of star clusters.
Consequently, the output at 100 Myr shows many bright, young star clusters with relatively concentrated sizes.
At 150 Myr, the star clusters become more disrupted and the number of star clusters has decreased.
This is due to tidal disruption and lack of resolution that leads to artificial disruption as in \citet{2024ApJ...974..193J}.
The decrease can be seen in the number of star clusters in the galaxy shown in Fig.~\ref{fig:host_properties}.

Fig.~\ref{fig:profile} exhibits the profiles of each component---dark matter ({\it green}), gas ({\it cyan}), stars outside the NSC ({\it orange}), stars in the NSC ({\it red})---of the host galaxy at $t=$0, 50, 100, and 150~Myr.
The initial condition of the NSC at 0 Myr follows the Plummer profile.
As the NSC evolves, it transitions into denser and cuspier profiles, whereas the profiles of dark matter, stars, and gas within the host galaxy remain consistent with the initial conditions.
The gravitational influence of the NSC results in the accretion of gas and dark matter into the NSC.
Owing to the constraints of resolution, the profiles of gas and dark matter are unable to extend completely to the center of the NSC.
\tr{However, the profile and dynamics of stars within a star cluster are accurately simulated due to the direct N-body integration.}
Subsequent to the merger occurring around 140 Myr, the cuspy profile of the NSC transitions into a core configuration by 150 Myr (refer to Fig.~\ref{fig:mass_growth_merger}). 
Nonetheless, the density of dark matter particles within the NSC remains preserved.
This hints that the dynamics of dark matter may have a substantial impact on the NSC.
However, the limitations in mass and force resolution for dark matter in this study impede a more detailed investigation of its effects, as elaborated in Sec.~\ref{sec:appendix_dark_matter}.

We also verify that the NSC remains dynamically stable at the galactic center throughout the simulation.
Fig.~\ref{fig:trajectory} shows the trajectory of the center of mass of the NSC over 150~Myr.
The initial position coincides with the center of mass of the galaxy ({\it blue} dot), while the final position at 150~Myr is drawn with the {\it red} dot.
The NSC never strays beyond $\sim 100\pc$ from the galactic center.
Although we only present the fiducial simulation here, the impact of \tr{supernova} feedback and star formation on the trajectory is minor across all runs.
In addition, our NSC remains the most massive star cluster in the galactic center throughout the simulation.
Consequently, owing to the relatively short travel distances and its mass, it is reasonable to consider the initially implanted NSC as the actual NSC of its host galaxy.

\subsubsection{Global Star Formation of the Host Galaxy}
\label{sec:result_global_sf}

We now focus on the star formation trends in the host galaxy.
Fig.~\ref{fig:host_properties} depicts the total mass of the newly formed stars ({\it top}), mass fraction of the total newly formed star clusters to the newly formed stars ({\it middle}), and the instantaneous number of newly formed star clusters over 150 Myr.
Here, `newly formed' denotes stars formed after the start of the simulations, excluding the stars in the initial conditions---the initial disk stars and bulge stars (see Sec.~\ref{sec:method_IC}).
\tr{The stars lose mass through supernova feedback by ejecting a fraction of their mass into the gas cells (see Sec.~\ref{sec:method_star_physics}).
Thus, the mass of newly formed stars denotes the sum over the instantaneous mass of newly formed stars in the galaxy at each time step.}
\tr{The feedback variation impacts the total mass of newly formed stars only negligibly, as seen by comparing `No Feedback' with `Fiducial'.}
However, \tr{since the stars in `No Feedback' do not lose mass through supernova feedback, the actual cumulative mass of newly formed stars at birth is greater in the other three simulations with supernova feedback than in `No Feedback'.}
\tr{More interestingly, even the simulation with twice the gas mass does not produce more stars than other simulations.}

In contrast, the star cluster populations exhibit clearly different patterns.
The mass fraction of star clusters is the ratio of the integrated mass of all the star clusters within the galaxy at each time to the total mass of newly formed stars.
Although the masses of newly formed stars are similar in the \tr{four} simulations ({\it top}), the mass of star clusters is higher in the fiducial simulation followed by the low star formation, double gas mass, and no feedback simulations, respectively ({\it middle}).
Yet, the numbers of star clusters in the galaxy ({\it bottom}) are more or less comparable across the \tr{four} simulations.
With this, we can deduce that the masses of star clusters in the no feedback simulation are relatively lighter than those in the fiducial simulation.
In other words, \tr{supernova} feedback reshapes the star cluster mass function: it suppresses the formation of low-mass clusters by disrupting small gas clumps, while concentrating gas into fewer, more massive structures that produce heavier star clusters (see also Fig.~\ref{fig:cluster_count}).
In the absence of this regulatory mechanism, the `No Feedback' run produces a proliferation of lighter clusters.
The number of the star clusters in the three simulations rapidly declines after the peak at 80 Myr. 
This phenomenon can be attributed to the limited force resolution in the outer region of the galaxy, resulting in destruction of star clusters after several tens of Myr, as the direct N-body region is only confined to a sphere centered at the center of mass of the NSC with a radius of 200~pc.

\tr{Lastly, the double gas mass simulation has produced the similar amount of stars compared to other simulations, yet forming relatively fewer star clusters despite its twice amount of gas mass in the galaxy.
This is attributed to bursty star formation in the first 5~Myr, which is not witnessed in other simulations (see the star formation rate panel of Fig.~\ref{fig:mass_growth_insitu}).
The early bursty star formation prevents further gas fragmentation and relatively dissipates the gas more than in the other simulations.
This results in the similar or even fewer formation of star clusters across the host galaxy.}


\begin{figure*}
    \includegraphics[width=0.97\linewidth]{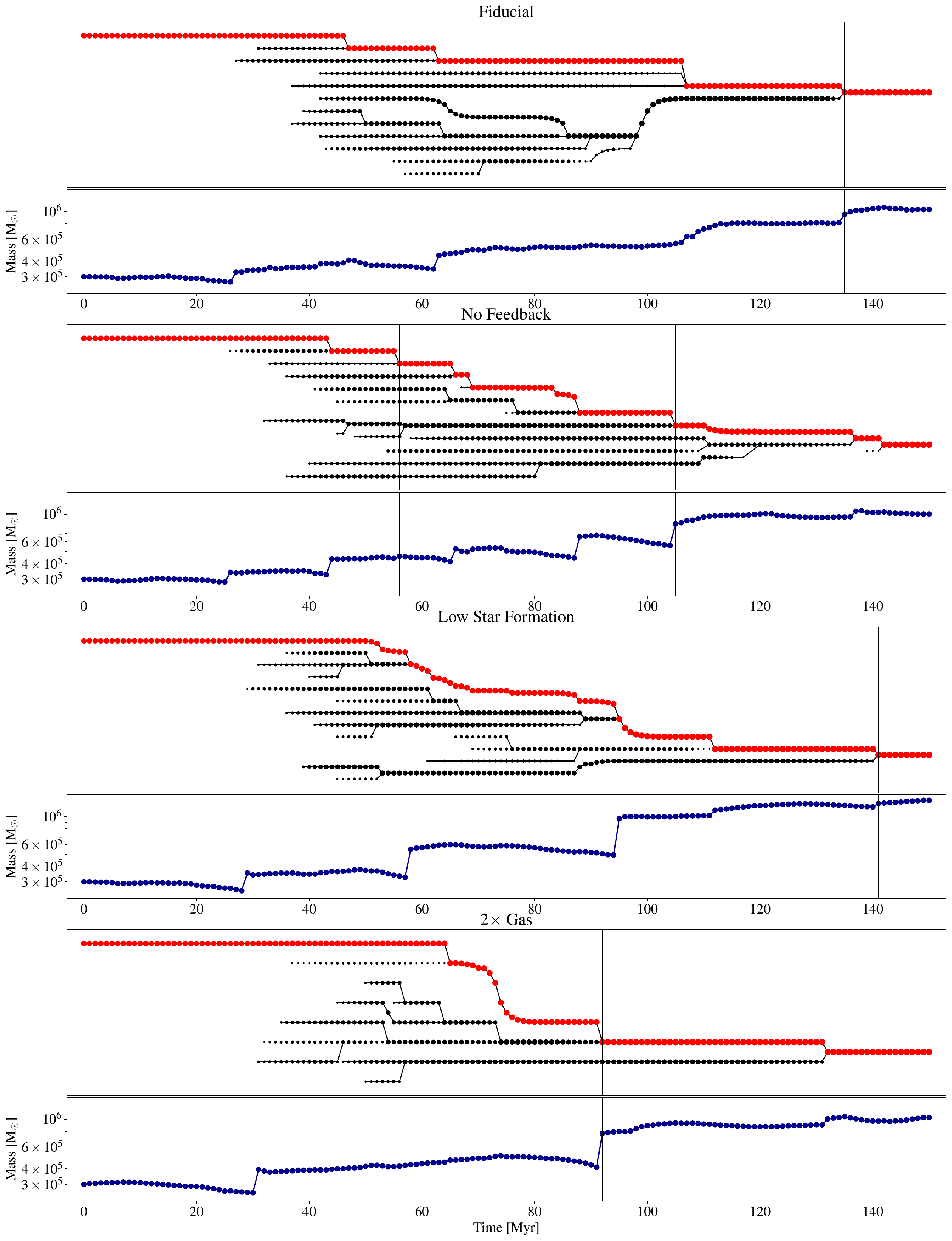}
    \caption{{\it Upper panels}: The merger trees and the mass growths of the NSCs for the three simulations---Fiducial ({\it 1st row}), No Feedback ({\it 2nd row}), Low Star Formation ({\it 3rd row}), \tr{and $2\times$ Gas ({\it 4th row}).}
    The NSCs---primary of merger trees---are represented by the {\it red} dots in the merger trees, while the {\it black} dots represent the nearby star clusters.
    The vertical lines indicate NSC-star cluster mergers.
    {\it Lower panels}: The stellar mass evolution of the NSCs given by the {\tt HDBSCAN} ({\it blue}).
    }
    \label{fig:merger_tree}
\end{figure*}

\begin{figure}
    \centering
    \includegraphics[width=1.0\linewidth]{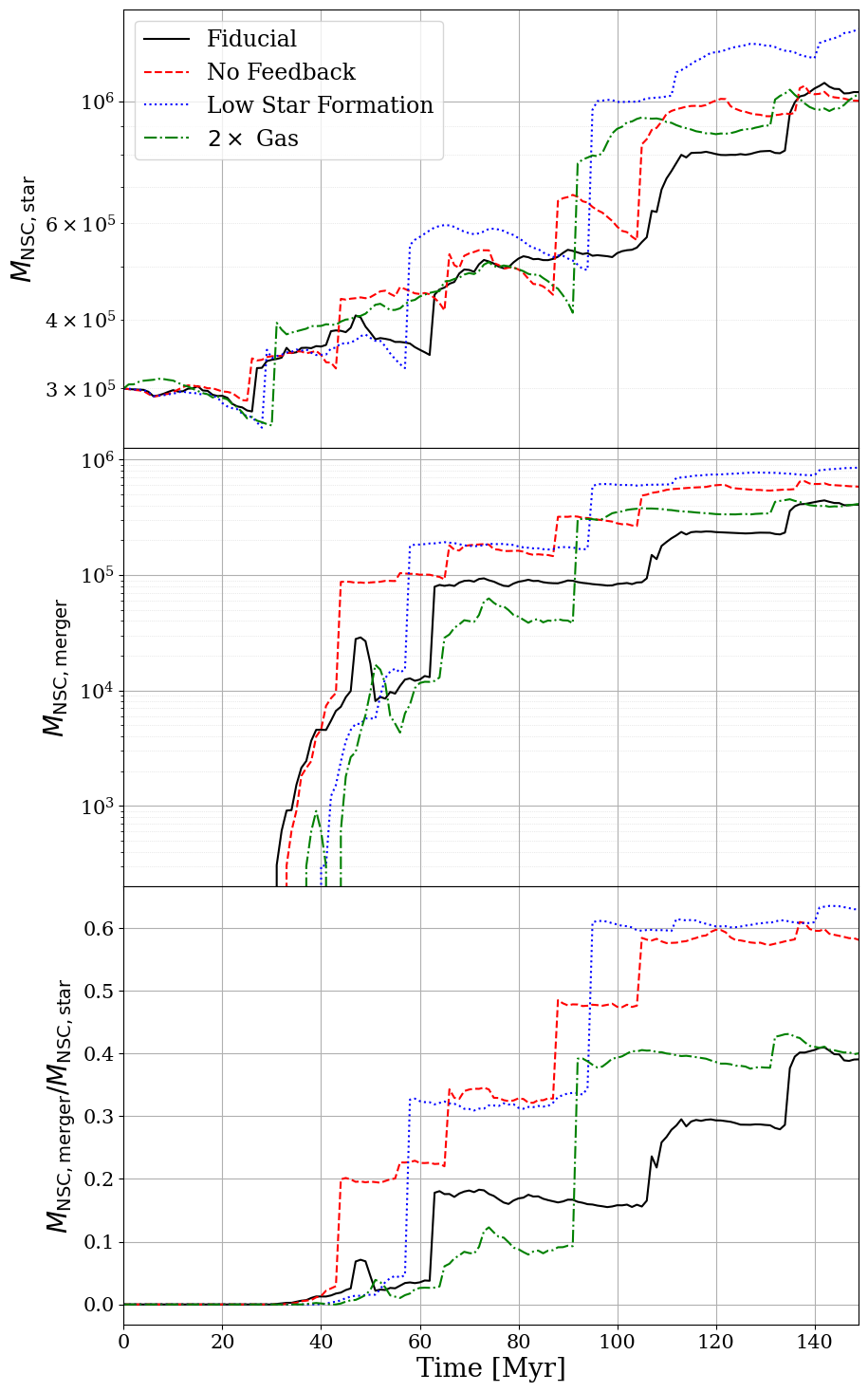}
    \caption{{\it Top}: the mass of the NSC over 150 Myr.
    {\it Middle}: the mass gain of the NSC from the merger with time.
    {\it Bottom}: the mass fraction of merger contribution to the total mass of the NSC versus time. 
   The fiducial, no feedback, low star formation simulations, \tr{and double gas mass} are represented by {\it black solid}, {\it red dashed}, {\it blue dotted}, and {\it green dot-dashed} lines, respectively.
   The discontinuities in the plots indicate the occurrence of mergers. 
   The merger contribution to the mass evolution of the NSC is accounted for by tracing the individual particles within the merging star clusters.
   }
    \label{fig:mass_growth_merger}
\end{figure}

\begin{figure}
    \centering
    \includegraphics[width=0.99\linewidth]{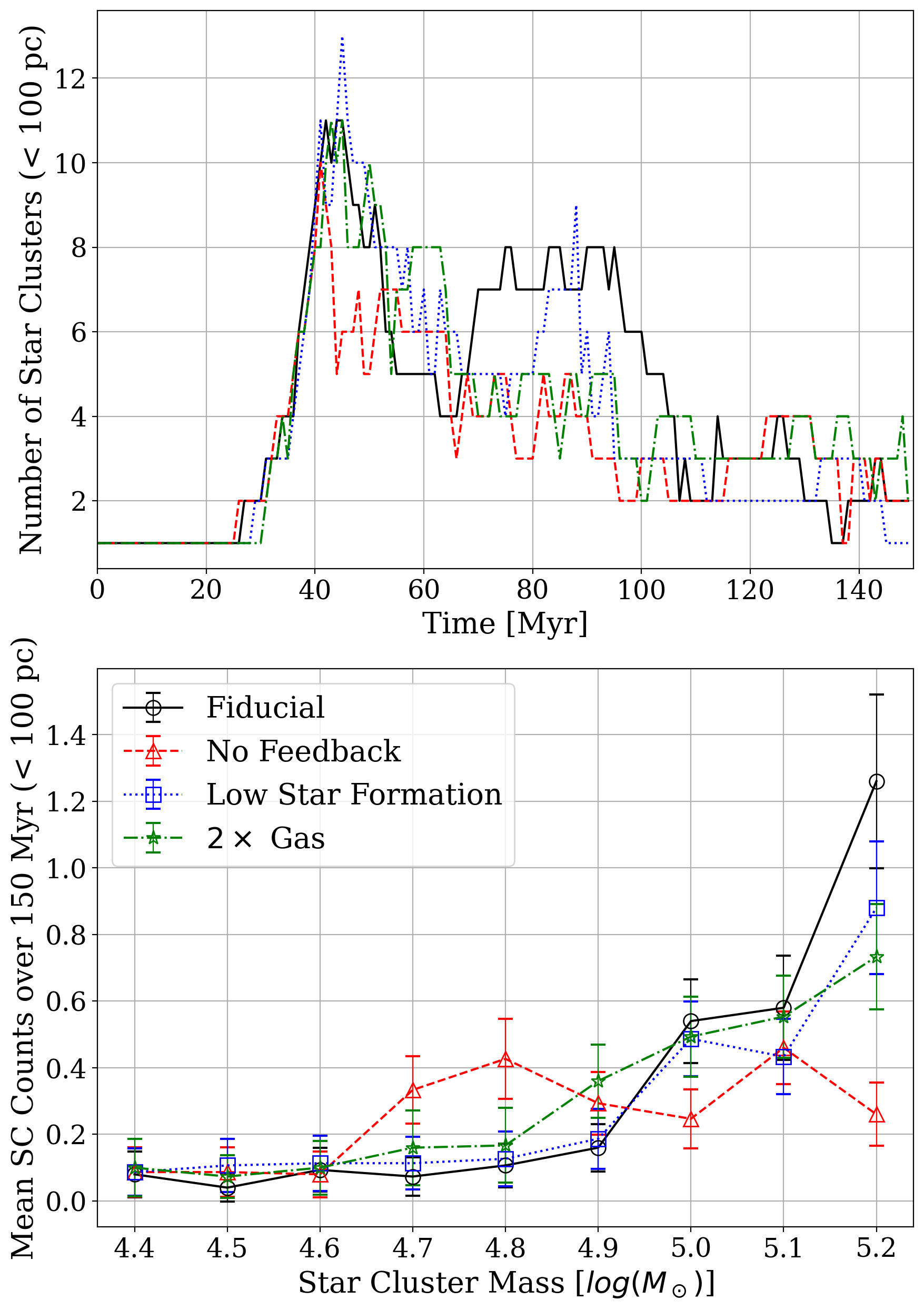}
    \caption{{\it Top}: The instantaneous number of the star clusters in vicinity of the NSC ($< 100 \pc$) as a function of time ({\it top}).
    {\it Bottom}: Histogram of mass of the star clusters averaged over 150 Myr.
    The `No feedback' run ({\it red dashed}) exhibits a reduced number of star clusters in the vicinity of the NSC compared to the `Fiducial' ({\it black solid}), `Low Star Formation' ({\it blue dotted}), \tr{and `$2\times$ Gas' ({\it green dot-dashed})} scenarios. 
    Notably, the `Fiducial' contains more massive star clusters, followed by the `Low Star Formation', while the `No feedback' shows a prevalence of smaller star clusters.}
    \label{fig:cluster_count}
\end{figure}

\subsection{Evolution via Mergers}
\label{sec:result_merger}


Fig.~\ref{fig:merger_tree} displays the merger trees and the mass evolutions of the NSCs from the `Fiducial' ({\it 1st row}), `No Feedback' ({\it 2nd row}), `Low Star Formation' ({\it 3rd row}), \tr{and `$2\times$ Gas' ({\it 4th row})} runs.
The progenitors---the NSCs---are represented by red dots.
Over the course of 150 Myr, there were \tr{five} mergers for both `Fiducial' and `Low Star Formation', while the `No Feedback' run experienced ten mergers.
However, the appreciable mass increases are fewer than the total number of mergers, suggesting that they are largely influenced by the major mergers characterized by a mass ratio of less than 5:1.
Provided that the jumps in the mass evolution are attributed to major mergers, `Fiducial', `No Feedback', and `Low Star Formation' have undergone three, five, and five major mergers, respectively, which is not proportional to the total number of mergers.
In the `Low Star Formation' run, the merger events occur at significantly high frequency, shortly after the formation of star clusters. 
\tr{The `$2\times$ Gas' run shows the least number of mergers (four mergers) among all the simulations, yet the mass of the NSCs at 150 Myr are comparable to the others.}
It is noteworthy that all simulations exhibit an initial spike that is not indicative of a merger. 
This phenomenon arises from algorithmic errors in the star cluster identification process, where the formation of the (first) nearby star clusters abruptly alters the criteria for overdensity, consequently leading to a reduced radius of the NSCs.
In the absence of these errors, a smooth, gradual growth of the NSCs is expected.

Fig.~\ref{fig:mass_growth_merger} depicts the evolution of the total mass of the NSCs ({\it top}), the mass attributed to mergers ({\it middle}), and the mass ratio of the NSC attributed to mergers ({\it bottom}).
The mass growth of the NSC shows significant jumps that coincide with merger events. 
The final total masses of the NSCs at 150 Myr are comparable across all \tr{four} runs despite the variations in stellar physics---$M_\mathrm{NSC,star}\sim10^6\msun$;
\tr{especially, except `Low Star Formation', the remaining three runs show very similar end masses.}
The `Low Star Formation' run attains the most mass growth, yet it exceeds the other simulations by only 20\%, amounting to $M_\mathrm{NSC,star}\sim1.23\times10^6\msun$ at 150 Myr.
The mass of the NSC attributed to mergers is determined by tracking the particles of the NSC that originate from other clusters, identified using the \texttt{HDBSCAN} method. 
In each snapshot, particles from other clusters are classified as merger contributions, and their aggregation provides the NSC mass contributions resulting from mergers. 
These masses originating from mergers ({\it middle}) have only mild differences, with the exception of the `Fiducial' run, which demonstrates the lowest value. 
In terms of fractions ({\it bottom}), the contribution of mergers is significantly lower by $\sim 50\%$ in the `Fiducial' run despite the fact that the final masses are comparable, indicating that other sources of mass growth are more predominant in the fiducial case.

Fig.~\ref{fig:cluster_count} illustrates the instantaneous number of star clusters in the vicinity of the NSC (within 100 pc) over time ({\it top}) and the histogram of mass of the star clusters averaged over 150 Myr ({\it bottom}).
The number of the star clusters refers to the count of the nearby star clusters within 100 pc around the NSC at each time step.
The `Fiducial' simulation ({\it black solid}) exhibits the highest abundance of star clusters, followed by the `Low Star Formation' ({\it blue dotted}), \tr{`$2\times$ Gas' ({\it green dot-dashed})}, and `No Feedback' ({\it red dashed}). 
Apart from a shared initial peak, these simulations do not exhibit any common trend.
The decline in abundance is consistent with the global trend of the host galaxy in Fig.~\ref{fig:host_properties}.
Intriguingly, the number of nearby star clusters has no consistent correlation with \tr{the frequency of} merger events, indicating that the detailed dynamics such as \tr{the balance between the conservation of angular momentum and dynamical friction-driven inward migration} are more important.

In the {\it bottom} panel, we average the histogram of mass of the nearby star clusters over 150~Myr.
The error bars represent the standard deviations of each mass bin, reduced by a factor of ten to enhance visual clarity.
Given the actual size of the standard deviations, the variance of each mass bin is considerably substantial.
The `No Feedback' run has significantly more light-mass star clusters of stellar mass from $10^{4.6}\msun$ to $10^{4.9}\msun$.
Meanwhile, `Fiducial', `Low Star Formation', and \tr{`$2\times$ Gas'} show similarity overall, except for the higher abundance at the high-mass end for the fiducial case.
The \tr{supernova} feedback suppresses the formation of small clusters in the `Fiducial', `Low Star Formation', and \tr{`$2\times$ Gas'} runs.
In addition, the `No Feedback' run shows more mergers---ten mergers---than the other simulations.
However, given similarity in the final masses, each mass of the merging clusters is relatively smaller.
\tr{Yet, the caveat is that the differences---in the number of mergers and cluster masses---are not significant enough to make a strong statement.}
The lighter star clusters are more susceptible to the tidal field of the NSC, which explains the higher merger frequency in the `No Feedback' simulation.
\tr{Notice that the star clusters within this region are well-resolved by the direct N-body solver and do not suffer from artificial dissipation or tidal forces.}

The observed merger timescales are consistent with theoretical expectations from dynamical friction.
The timescale of dynamical friction is
\begin{equation}
\begin{split}
    t_\mathrm{fric} &=  \frac{v_\mathrm{circ}r^2}{GM\ln\Lambda}\\
    &\approx \frac{10^2}{\ln\Lambda} \left(\frac{r}{1\kpc}\right)^2  \left(\frac{v_\mathrm{circ}}{200 \km/\s}\right) \left(\frac{10^5\msun}{M_\mathrm{SC}}\right)\mathrm{Gyr},
\end{split}
\end{equation}
where $v_\mathrm{circ}$, $r$, and $\ln\Lambda$ are the circular velocity ($\sqrt{GM_\mathrm{host}/r}$), the orbital radius of a star cluster, and the Coulomb logarithm (typically 10 to 20; \citealt{binney1987gady.book.....B}), respectively.
The mean mass of young star clusters that form in the simulations is approximately $10^5\msun$, never exceeding $10^6\msun$ (see Fig.~\ref{fig:cluster_count}), in agreement with \citet{portegies2010ARA&A..48..431P}.
The total mass of the host galaxy within $200\pc$ is approximately $3\times10^7\msun$ averaged over 150~Myr.
Consequently, the infall timescale of $\sim 10^5\msun$ star clusters from the outskirts is approximately $\sim 20$--$40\Myr$.
This is consistent with the merger rates of $\sim 1/20$--$1/40\;\mathrm{Myr}^{-1}$ observed in Fig.~\ref{fig:merger_tree}.


\begin{figure}
    \centering
    \includegraphics[width=0.99\linewidth]{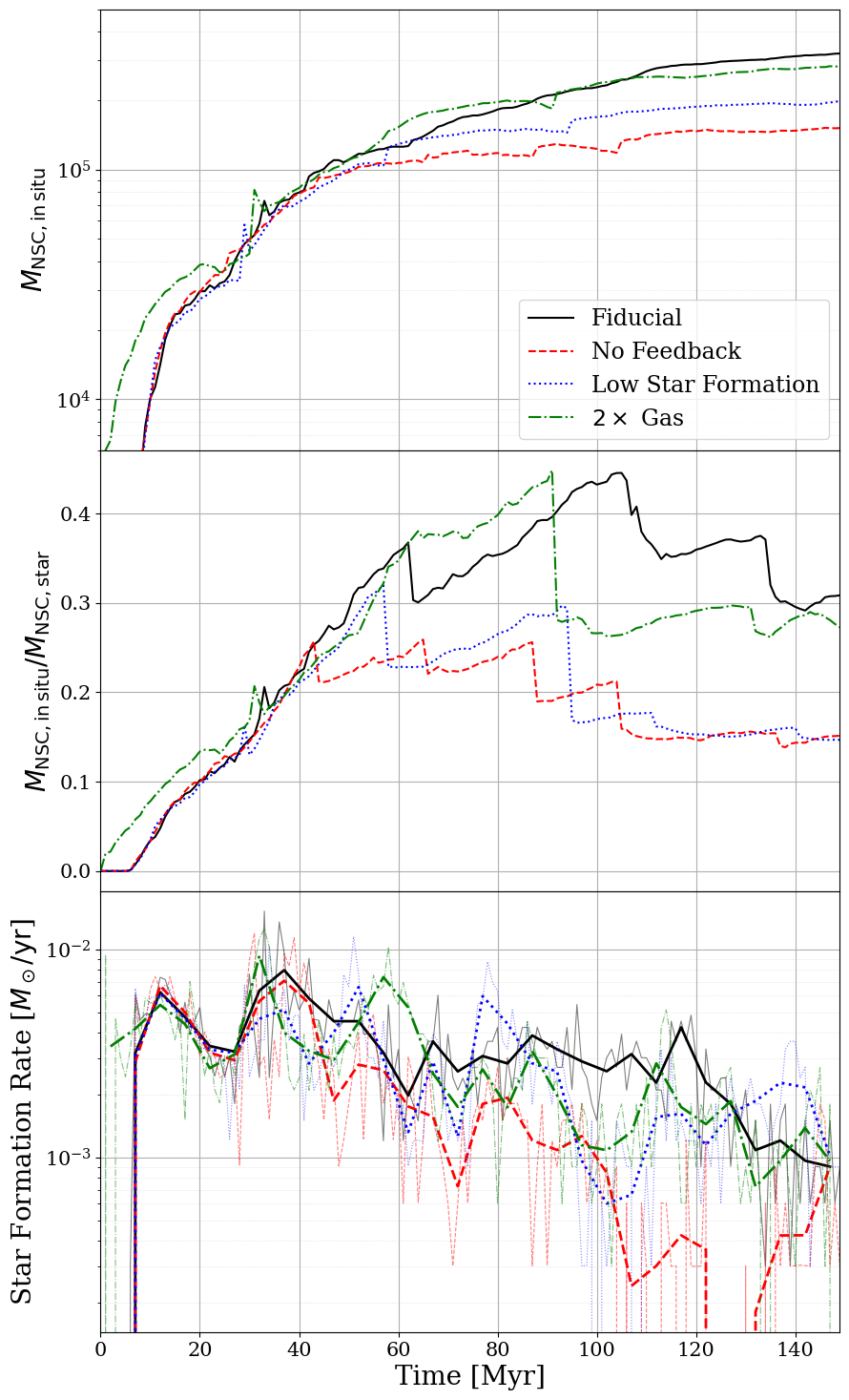}
    \caption{{\it Top}: The cumulative mass of stars formed inside the NSCs versus time. 
    {\it Middle}: The fraction of cumulative in-situ formed stellar mass versus the total mass of the NSCs with time.
    {\it Bottom}: Star formation rate within the NSCs versus time. 
    }
    \label{fig:mass_growth_insitu}
\end{figure}

\begin{figure}
    \centering
    \includegraphics[width=0.99\linewidth]{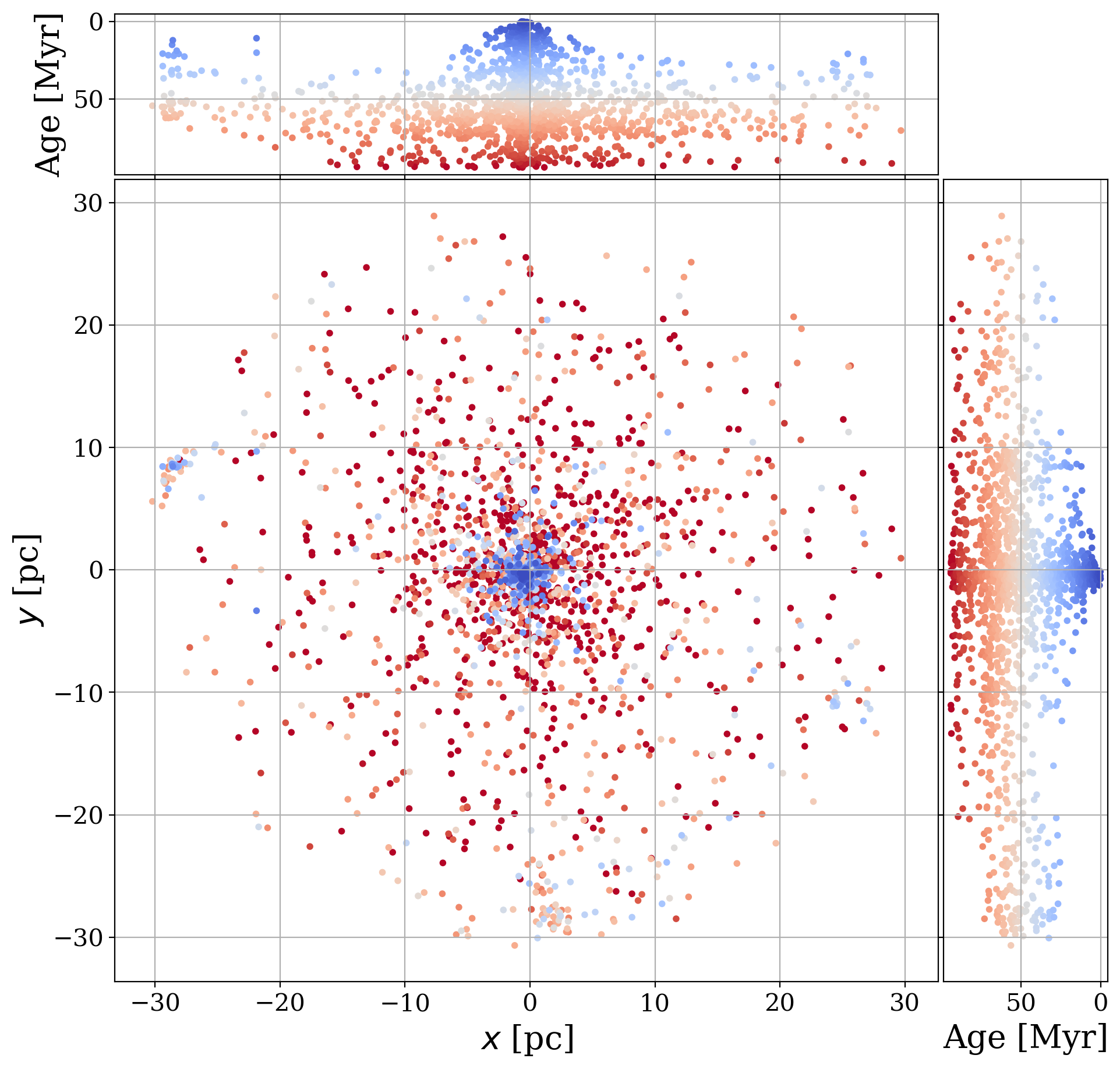}
    \caption{Spatial distribution of star particles\tr{---newly born after the simulations have begun---}in the NSC color-coded with the ages at 100 Myr.
    The bluer indicates a more younger star, whereas the redder suggests an older population ({\it top} and {\it right}).
    The central region exhibits a higher concentration of the younger population with age of $<20$ Myr.}
    \label{fig:spatial_age}
\end{figure}

\subsection{Evolution via In-Situ Star Formation}
\label{sec:result_insitu}

Having examined the contribution of mergers, we now turn to the competing growth channel: in-situ star formation within the NSC.
This process occurs when gas flows into the galaxy's center, cools, and collapses, leading to the formation of new stars.


Fig.~\ref{fig:mass_growth_insitu} shows the in-situ contribution to the growth of the NSC.
The {\it top} panel displays the aggregated masses of the star particles formed within the NSCs.
The ratios of the in-situ mass contribution to the total mass are shown in the {\it middle} panel.
In-situ star formation makes the largest contribution to the evolution of the NSC in the `Fiducial' ({\it black solid}), followed \tr{by `$2\times$ Gas'}, compared to the other two simulations.
This is consistent with the merger analysis in Sec.~\ref{sec:result_merger}: the `Fiducial' and `$2\times$ Gas' cases showed the lowest merger contribution despite having comparable final masses, implying that other sources of mass growth---namely in-situ star formation---are predominant.
\tr{The {\it bottom} panel illustrates the star formation rates averaged over 5~Myr ({\it thick} lines) and 1~Myr ({\it thin translucent} lines).}
The star formation rate in the fiducial run is the most active on average over 150 Myr compared to the `No Feedback' and `Low Star Formation' runs.
\tr{The run with double gas mass forms stars rapidly in the beginning of the simulation before 60~Myr, with minimal star formation afterward.}
\tr{The trends of in-situ star formation in the different simulations are related to the density of star cluster cores where most of the star formation occurs, which is also closely correlated with the merger events; we discuss this connection in detail in Sec.~\ref{sec:discussion_merger_regulates}.}

Fig.~\ref{fig:spatial_age} shows the spatial distribution of the NSC stars\tr{---newly born after the simulations have begun---}color-coded with their stellar ages at $100\Myr$.
A young population, with ages less than 20~Myr, is significantly concentrated within the central few parsecs, while the older stars are distributed relatively uniformly across the NSC.
However, due to the relatively short duration of the simulations ($\sim 150\Myr$), the stars are predominantly young.
Nevertheless, this is consistent with the findings of \citet{georgiev2014MNRAS.441.3570G, carson2015AJ....149..170C}, which identified a higher concentration of young stellar populations in the core regions of NSCs, and is further corroborated by \citet{fahrion2024A&A...687A..83F}, who identified a concentration of cold molecular $\mathrm{H}_2$ gas at the center.

Two notable features deserve comment, though we defer their full interpretation to the Discussion (Sec.~\ref{sec:discussion_interplay}).
First, it is counterintuitive that the fiducial simulation---with \tr{supernova} feedback and more efficient star formation---forms more stars in the NSC than the other runs.
Second, the in-situ star formation rate shows negligible correlation with merger events: unlike galaxy mergers, the star cluster mergers in our simulations do not trigger enhanced star formation.


\begin{figure*}
    \includegraphics[width=0.95\linewidth]{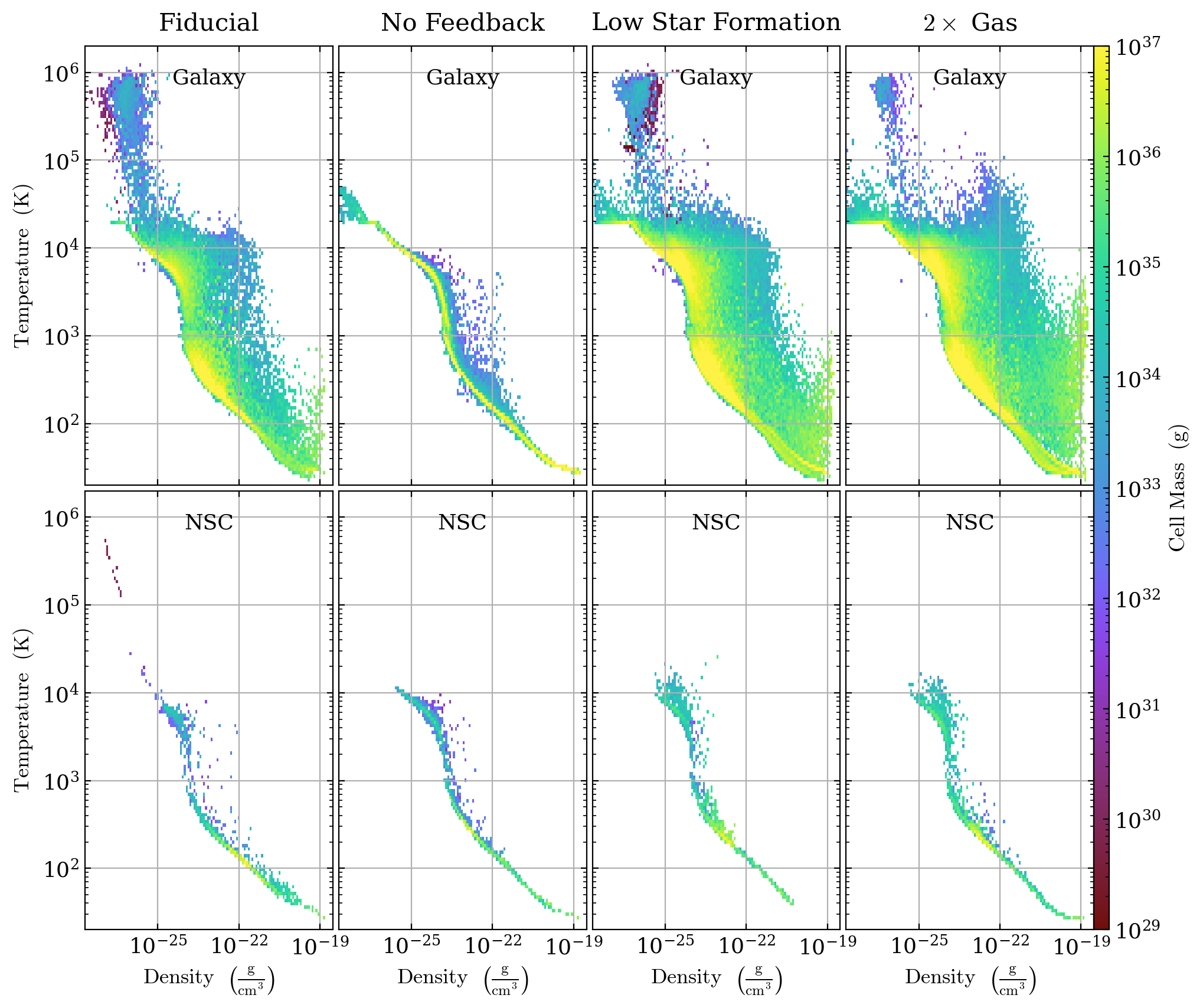}
    \caption{Phase plot of gas temperature, gas density, and gas (cell) mass of the host galaxy ({\it top}) and the NSC ({\it bottom}) at $t=100$ Myr.
    In the galaxies, the `Fiducial' (left) and `Low Star Formation' (right) show hot, diffuse gas in the upper-left corner, indicating supernova feedback from stars, which is absent in the `No Feedback' ({\it middle}). 
    On the other hand, the gas properties within the NSCs are similar regardless of star formation and \tr{supernova} feedback.
    }
    \label{fig:phase}
\end{figure*}

\begin{figure}
    \includegraphics[width=0.98\linewidth]{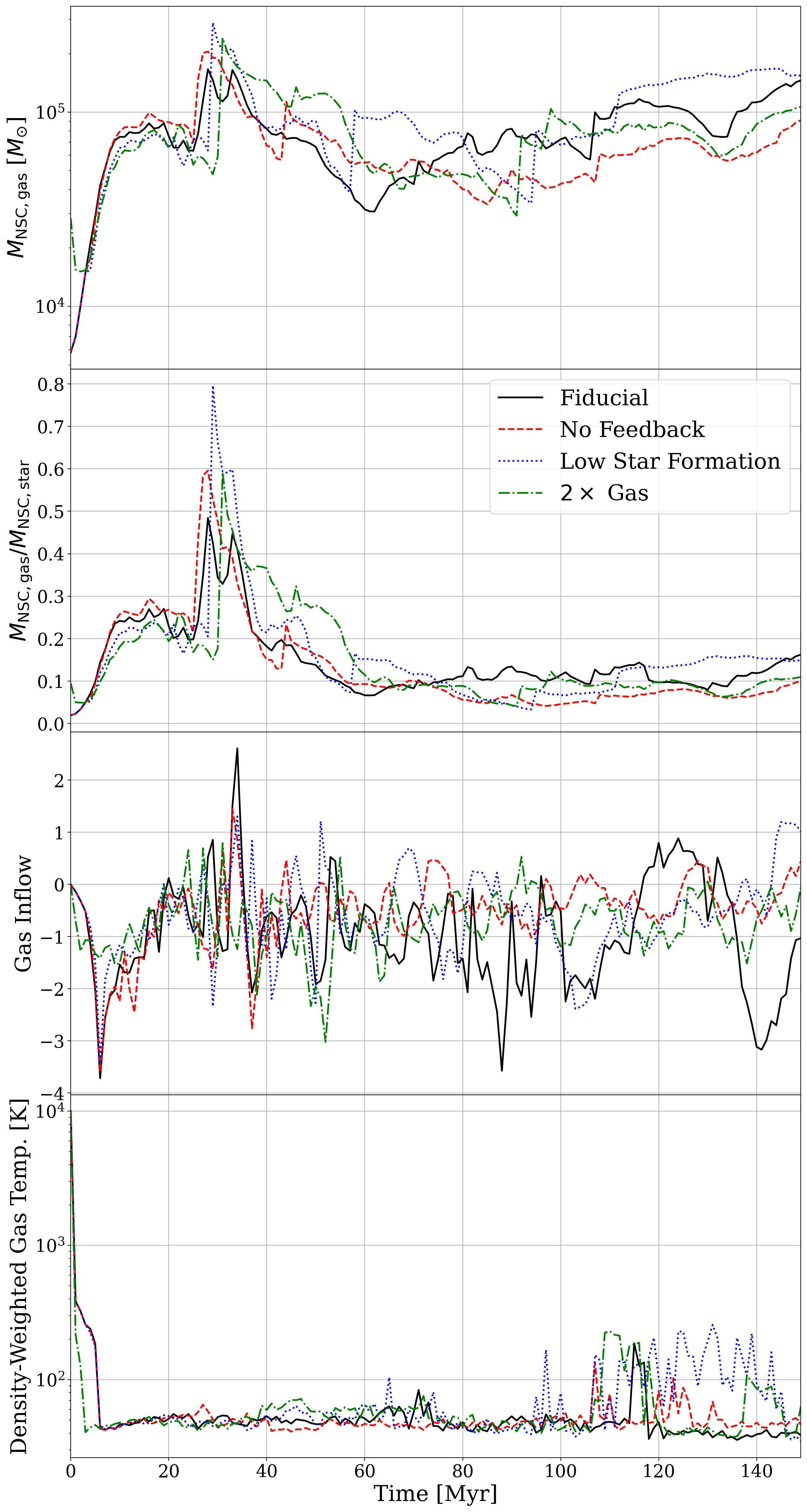}
    \caption{Evolution of gas mass ({\it first row}), mass fraction of gas compared to stars ({\it second row}), gas inflow ({\it third row}), and gas density-weighted mean gas temperature ({\it fourth row}) within the NSCs.
    \tr{In the gas inflow ({\it third row}), a negative value represents the infall of gas towards the center.}}
    \label{fig:gas_properties}
\end{figure}

\subsection{Gas in the Nuclear Star Cluster}
\label{sec:result_nsc_gas}


The gas content of the NSC is intimately linked to its ability to sustain in-situ star formation.
Here we examine the gas properties within the NSC across all four simulations.

The key finding is that, despite marked differences in the host galaxy gas phases, the gas properties within the NSCs are similar across all runs.
Fig.~\ref{fig:phase} shows the galaxy-wide ({\it top}) and NSC-wide ({\it bottom}) phase plots of gas temperature, gas density, and cell mass at 100~Myr.
At the galaxy scale, the `Fiducial' ({\it 1st column}), `Low Star Formation' ({\it 3rd column}), and `$2\times$ Gas' ({\it 4th column}) exhibit regions of hot, diffuse gas in the {\it upper-left} corner, indicative of supernova feedback; this signature is absent in the `No Feedback' run ({\it 2nd column}).
The `Low Star Formation' \tr{and `$2\times$ Gas'} runs retain more dense gas cells ({\it lower-right} corner) due to the lower star formation efficiency and the greater gas content, respectively.
In the absence of \tr{supernova} feedback, the `No Feedback' run shows a relatively simple gas phase that adheres to the attributes of radiative cooling and heating.
However, the gas phases within the NSCs ({\it bottom row}) are strikingly uniform across all simulations.
\tr{Within the NSCs, the trail of supernova feedback---hot diffuse gas in the upper-left corner---only appears in the `Fiducial' run, but its impact on NSC growth is negligible: the total mass of the hot diffuse gas is less than $1\msun$, insignificant compared to the NSC mass and star formation rates shown in Figs.~\ref{fig:mass_growth_merger} and \ref{fig:mass_growth_insitu}.}
This pattern persists consistently over the duration of the 150~Myr run, even during mergers.

\tr{That being said, supernova feedback is not entirely inconsequential within NSCs.
The thermal energy injected by supernovae heats the gas and intermittently suppresses star formation, but not persistently or consistently.
This is because the dense nature of NSCs gravitationally overwhelms the gas pressure resulting from supernova heating.
However, the spatial resolution of the gas cells limits investigation of the detailed gas dynamics inside the NSCs, and the results largely depend on the adopted subgrid models.
Further discussion on these limitations is provided in Sec.~\ref{sec:discussion_limitations}.
Lastly, the phase plot is generally determined by cooling functions and subgrid models.
A comprehensive comparison to different simulations can be found in Fig.~17 of \citet[][]{agora2016ApJ...833..202K}.}

Fig.~\ref{fig:gas_properties} displays the evolution of gas mass ({\it first row}), mass fraction of gas compared to stars ({\it second row}), gas inflow ({\it third row}), and averaged gas temperature ({\it fourth row}) within the NSCs. 
In contrast to the stellar mass of the NSCs, the gas mass remains relatively constant at $\sim10^5\msun$ over 150~Myr, except for the initial inflow and merger events.
\tr{The large initial inflow is driven by an imbalance between angular momentum of gas and gravity in the vicinity and inside the NSCs.
In the initial conditions, star clusters do not contain any gas so the circular velocity of gas within the NSC region is not initialized with the rotational velocity of the star cluster, but matches that of the host galaxy which is lower than that of the NSC. This triggers a dominant infall of gas due to gravity of the NSC (refer to Sec.~\ref{sec:discussion_limitations} for details).}
The gas mass fraction rapidly grows in the first 30~Myr and gradually decreases over time, settling at $\sim10-20\%$ across all simulations.
In the fiducial simulation, the increased gas inflows may facilitate enhanced star formation, notwithstanding the stable gas mass, indicating that these systems retain a certain quantity of gas with any surplus being converted into stars.
\tr{However, there is no signature that mergers effectively bring a noticeable amount of gas into the NSCs.}
The density-weighted mean temperature of the gas remains low, not exceeding 100~K on average, indicating that cooling is highly effective within the NSCs \citep{2020ApJ...897..176K}.


\subsection{Rotational Dynamics of NSCs}
\label{sec:result_rotation}

\begin{figure}
    \centering
    \includegraphics[width=1.0\linewidth]{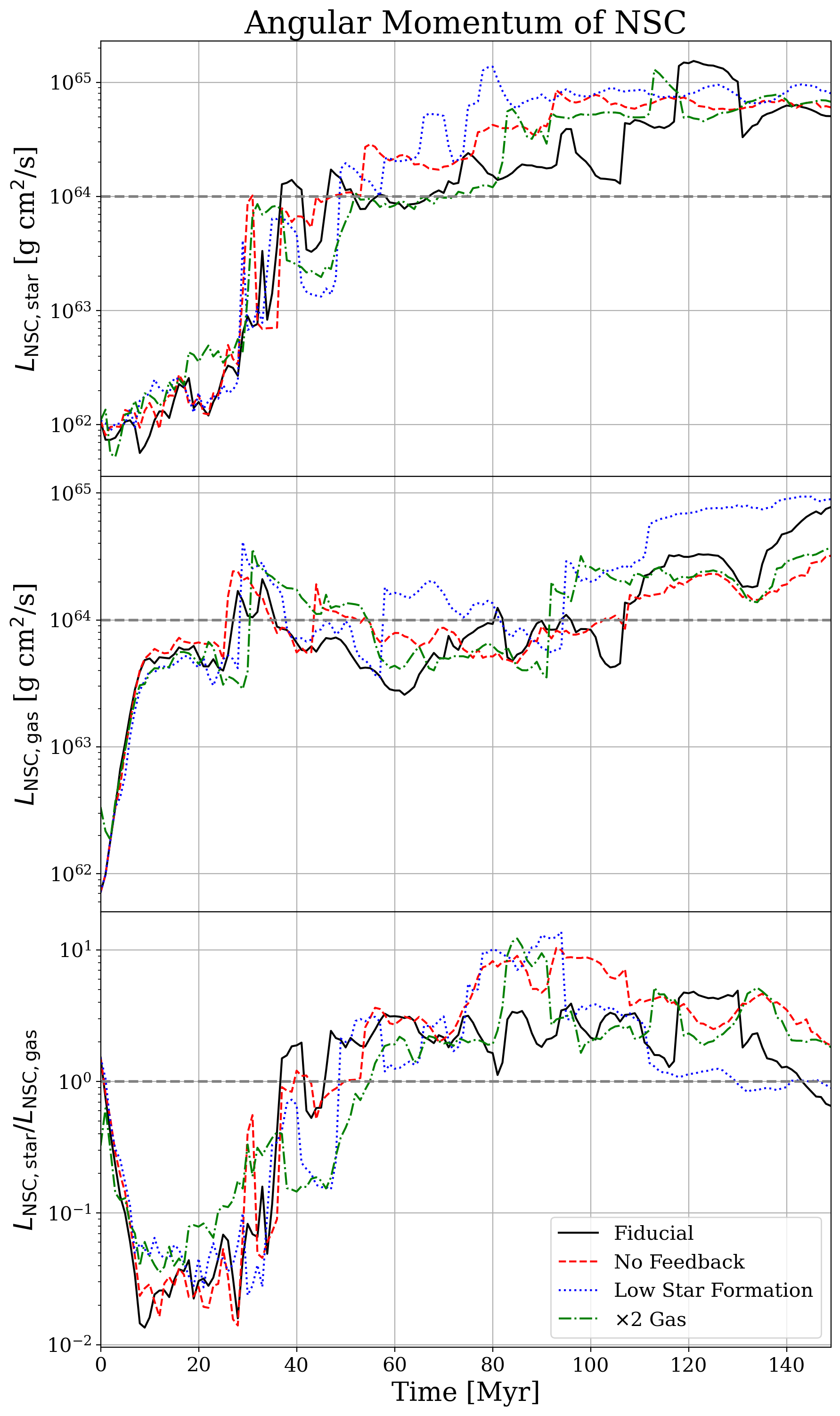}
    \caption{Time evolution of the angular momentum in the NSCs.
             {\it Top}: The angular momentum of NSC stars. 
             {\it Middle}: The angular momentum of NSC gas.
             {\it Bottom}: The ratio of stellar to gas angular momentum.}
    \label{fig:angular_momentum}
\end{figure}

\begin{figure*}
    \centering
    \includegraphics[width=1.0\linewidth]{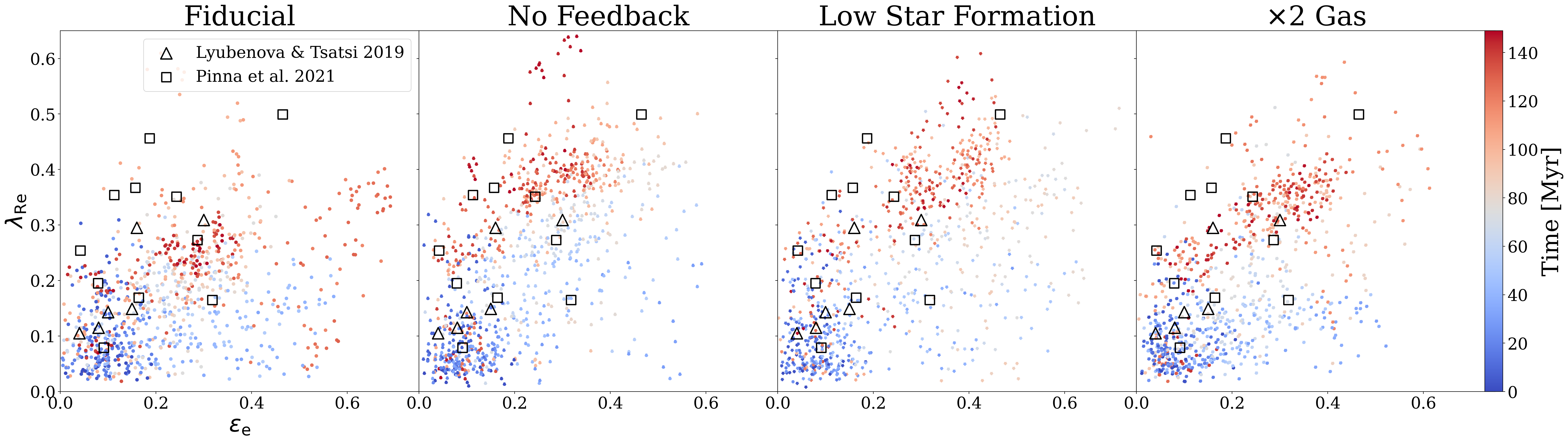}
    \caption{\tr{Distribution of ellipticity ($\varepsilon_e$) versus specific angular momentum ($\lambda_{R_e}$) within one effective radius for simulated NSCs under four different scenarios: Fiducial, No Feedback, Low Star Formation, and $2\times$ Gas. 
    Each data point represents a measurement at different epochs, color-coded according to simulation time (0--150 Myr). 
    Observational data from \citep[][{\it black triangles}]{2019A&A...629A..44L} and \citet[][{\it black squares}]{pinna2021ApJ...921....8P} are overlaid for comparison.}}
    \label{fig:lam_e-eps}
\end{figure*}

\tr{Fig.~\ref{fig:angular_momentum} shows the time evolution of angular momentum in the NSCs for all four simulation setups.
The {\it top} panel displays the stellar component, which initially starts at $\sim10^{62}\g\cm^2\s^{-1}$ and rises steeply between 20 and 40~Myr as the central region is assembled through gas inflow and early in-situ star formation. 
After 40~Myr the stellar angular momentum continues to increase on average, reaching $\sim10^{65}\g\cm^2\s^{-1}$ by the end of the simulation. 
The `Low Star Formation', `No Feedback', and `$2\times$ Gas' runs exhibit angular momentum gains during mergers---suggesting well-aligned, coplanar interactions---whereas the `Fiducial' run does not, implying that its mergers are more randomly oriented.}

\tr{The {\it middle} panel traces the gas angular momentum, which exhibits a rapid rise by $\sim20$~Myr driven by the initial gas inflow onto the originally gas-free NSCs.
The ratio of stellar to gas angular momentum, shown in the {\it bottom} panel, remains far below unity until about 40~Myr, indicating that the gas predominantly carries the NSC's angular momentum in the early phases. 
Between 40 and 60~Myr, gas and stellar angular momenta converge near $\sim10^{64}\,\mathrm{g\,cm^2\,s^{-1}}$, reflecting a phase where stars form within a coherently rotating gas reservoir---consistent with the comparable in-situ star formation rates across runs at that epoch (see Fig.~\ref{fig:mass_growth_insitu}).}

\tr{To compare with observations, we project our NSCs onto the ellipticity--specific angular momentum plane.
The ellipticity is defined as
\begin{equation}
\varepsilon_e = 1 - \frac{b}{a},
\end{equation}
where $a$ and $b$ represent the semi-major and semi-minor axes of the projected stellar distribution.
The specific angular momentum parameter $\lambda_{R_e}$ measures the importance of rotation relative to velocity dispersion within one effective radius~$R_e$:
\begin{equation}
\lambda_{R_e} = \frac{\sum_{i} F_i R_i |V_i|}{\sum_{i} F_i R_i \sqrt{V_i^2 + \sigma_i^2}},
\end{equation}
where $F_i$, $R_i$, $V_i$, and $\sigma_i$ are the flux (or mass weight), radius, line-of-sight velocity, and velocity dispersion of the $i$-th spatial bin, respectively.}

\tr{Fig.~\ref{fig:lam_e-eps} presents the $\varepsilon_e$--$\lambda_{R_e}$ distribution for the four runs, with each point corresponding to a particular epoch from 0 to 150~Myr.
To reduce line-of-sight effects, we project the NSCs onto five random planes and overlay each projection in its panel.
Observational data from \citet{2019A&A...629A..44L} and \citet{pinna2021ApJ...921....8P} are plotted as reference points.
All simulations show a broadly similar trend of evolving toward higher $\varepsilon_e$ and $\lambda_{R_e}$, implying increased flattening and rotation over time.
The `No Feedback' and `$2\times$ Gas' runs exhibit particularly noticeable shifts toward higher $\lambda_{R_e}$, suggesting a stronger rotational influence.
Observational data points overlap substantially with the simulated parameter space, supporting the realism of the NSC kinematics captured by our models.
However, discerning precise evolutionary pathways remains challenging due to substantial scatter, and our non-individual star particles may amplify fluctuations beyond their intrinsic levels (see Sec.~\ref{sec:discussion_limitations}).
The physical interpretation of these rotational trends is discussed in Sec.~\ref{sec:discussion_degeneracies}.}


\section{Discussion}
\label{sec:discussion}


\subsection{Interplay of Mergers, In-Situ Star Formation, and Feedback}
\label{sec:discussion_interplay}

\subsubsection{Competing Growth Channels}
\label{sec:discussion_channels}

\begin{figure}
    \centering
    \includegraphics[width=1.0\linewidth]{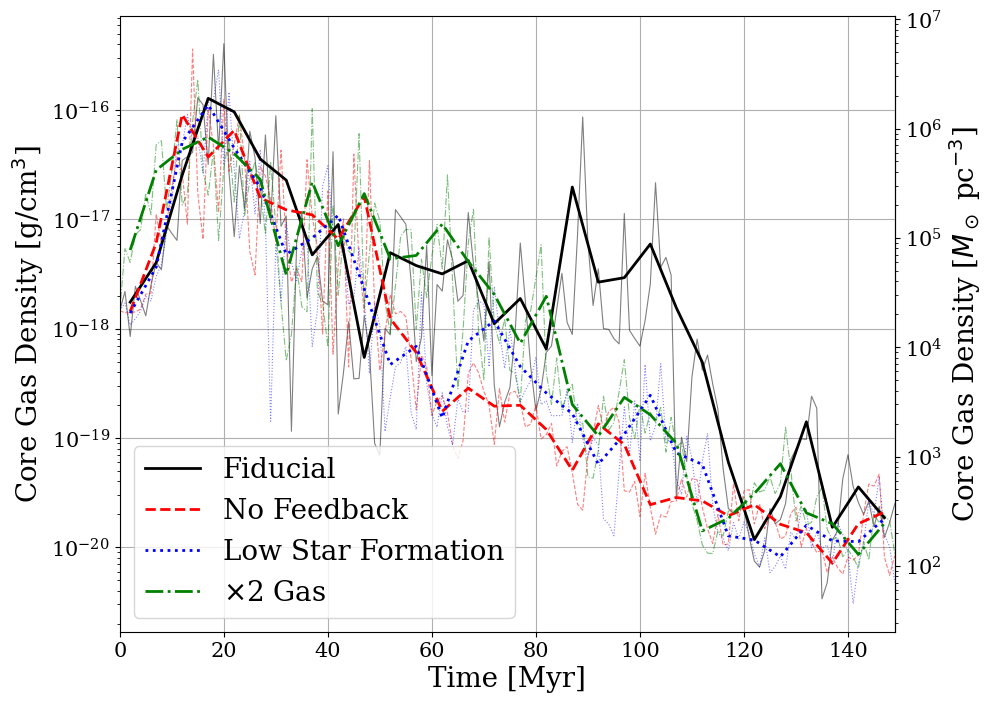}
    \caption{Core density of gas within NSCs with time.
    The volume within which the core density is calculated is determined by the volume that 20 nearest particles from the center of mass of the NSCs occupy.
    \tr{The {\it thick} line represents the core density averaged over 5 Myr, and the {\it thin translucent} line shows the instantaneous core density.}}
    \label{fig:core_density}
\end{figure}

The inward migration of star clusters plays a key role in the growth of NSCs. 
Dynamical friction causes these star clusters to spiral toward the galaxy's center, where they either merge with, or are tidally disrupted by, the NSC. 
Throughout the simulations detailed in this study, mergers have been a significant factor in the mass growth of the NSC, particularly during major merger events (see Sec.~\ref{sec:result_merger}).
While the total number of mergers varied across different simulations, their contribution to the overall mass of the NSC is not always proportional to their frequency.
By the end of the simulations at 150~Myr, the masses of the NSCs are comparable, suggesting that there exists an inverse proportionality between the frequency of mergers and the mass of the secondary star clusters (see Fig.~\ref{fig:mass_growth_merger}).

On the other hand, gas funneled into the center of the galaxies and the NSCs can cool and condense, leading to episodes of star formation. 
This mechanism is responsible for producing young stars within the cluster (see Fig.~\ref{fig:spatial_age}). 
It is believed that supernova feedback helps regulate star formation by preventing excessive gas accumulation and disrupting low-mass gas clumps, thereby reshaping the cluster mass function \citep{2019MNRAS.488.4753D,2024A&A...681A..28A}. 
However, in-situ star formation is more active in simulations with feedback and enhanced gas inflow on average, contributing significantly to the NSC's mass (see Fig.~\ref{fig:mass_growth_insitu}).

In the `Fiducial' model and `$2\times$ Gas' case, mergers and in-situ star formation contribute approximately equally to the NSC mass growth, with mergers slightly dominating.
In the `No Feedback' and `Low Star Formation' scenarios, mergers clearly dominate, accounting for more than 50\% of the mass increase, whereas in-situ star formation contributes less than 20\%.
The remaining $\sim 30$\% is composed of stars initially present or randomly accreted from the surrounding environment.
\tr{As a result, NSCs in our configuration show highly self-regulatory trends so that all four runs produce comparable NSC masses in a fashion insensitive to the variations of supernova feedback, star formation threshold, and gas content in the host galaxy.}

\subsubsection{Mergers Regulate In-Situ Star Formation}
\label{sec:discussion_merger_regulates}

There are two primary sources of gas inflow onto the NSC: inflow from the surrounding environment and gas delivered by merging clusters. 
However, the contribution of gas from mergers is often negligible because the gas fraction within star clusters is relatively consistent, thus not substantially altering the gas content within the NSC. 
Moreover, the mergers between NSCs and star clusters in our simulations are typically quiescent, proceeding through tidal disruption rather than violent head-on collisions; consequently, there is no shock-induced star formation.
Furthermore, tidal heating during mergers can hinder subsequent gas inflow into the NSC.

\tr{To investigate the link between mergers and in-situ star formation more closely, we examine how merger episodes influence the core gas density of the NSC---the key indicator of in-situ star-forming activity.
Fig.~\ref{fig:core_density} traces the evolution of the gas density in the central region of the NSCs over 150~Myr.
Here, the core gas density is defined as the volume density of gas contained within a sphere whose radius equals the distance to the 20th nearest star particle from the center of mass of the NSC.
On average, the core gas density and in-situ star formation rate are tightly correlated (cf.\ the star formation rate panel of Fig.~\ref{fig:mass_growth_insitu}): most star formation clearly occurs within NSC cores where gas densities are highest (Fig.~\ref{fig:spatial_age}).}

\tr{The comparison between core gas density evolution and merger events reveals that mergers systematically dissipate central gas, thereby suppressing in-situ star formation.
For instance, in the `$2\times$ Gas' run, the merger at $\sim 90$~Myr sharply lowers the core gas density, while the `Fiducial' run continues to build central gas until $\sim 120$~Myr.
These density dips precede the merger timestamps in Figs.~\ref{fig:merger_tree} and \ref{fig:mass_growth_merger}, since actual mergers span $\sim 10$~Myr (see Sec.~\ref{sec:method_hdbscan}) and the merging clusters experience tidal forces before formal coalescence.}

\tr{The contrast is stark: the `Fiducial' and `$2\times$ Gas' runs accumulate dense core gas and sustain higher in-situ star formation, whereas the `No Feedback' and `Low Star Formation' runs---characterized by more frequent mergers---exhibit repeated core gas disruptions and suppressed star formation.
This result is counterintuitive compared to typical merger-induced star formation scenarios in galaxy mergers, which rely on tidal compression to drive bursts of star formation \citep[e.g.,][]{2020ApJ...899...25S,2022MNRAS.516.4922R}.
In our simulations, however, tidal compression does not occur, and this may be due to one or more of the following factors:
(1) these are dry (gas-poor) cluster mergers, not the wet mergers where gas inflows fuel starbursts (see Sec.~\ref{sec:result_nsc_gas});
(2) the relative speed of the merging clusters is too low to generate the impulsive tidal forces required to compress gas and trigger star formation;
(3) our spatial resolution ($0.625\pc$ at finest) is relatively coarse compared to the internal scales of star clusters ($<10\pc$), likely under-resolving any compressive tidal effects.
In short, in our models mergers regulate in-situ star formation by disrupting the dense core gas reservoir.
To capture the detailed gas dynamics---especially tidal compression and shock-induced star formation---higher-resolution simulations with various configurations are required.}

\subsubsection{Self-Regulation and Degeneracies}
\label{sec:discussion_degeneracies}

\tr{In our simulations, no straightforward correlation emerges between NSC growth and the three physical variations. 
Supernova feedback reshapes the star cluster mass function across the host galaxy: by disrupting small gas clumps, it suppresses the formation of low-mass clusters while concentrating gas into fewer but more massive structures (see Figs.~\ref{fig:host_properties} and \ref{fig:cluster_count}).
Yet, this restructuring does not impede the growth of the NSC: the final NSC masses in the `Fiducial' and `No Feedback' runs remain comparable.
In the `No Feedback' run, the absence of feedback leads to a proliferation of lighter star clusters in the vicinity of the NSC (see Fig.~\ref{fig:cluster_count}), which---being more susceptible to tidal capture---promotes more frequent but lower-mass mergers.
Meanwhile, in the `Fiducial' run, enhanced in-situ star formation becomes the dominant contributor to NSC mass growth.}

\tr{In the `Low Star Formation' case, the formation of star clusters is generally suppressed both across the host galaxy and around the NSC.
Interestingly, however, we observe several spikes in the star cluster population in the vicinity of the NSC, particularly before the major merger events (see Fig.~\ref{fig:cluster_count}).
These fluctuations may arise from stochasticity rather than underlying astrophysics.
In our current setup, it is difficult to distinguish between genuine physical effects and random noise.
Future work should clarify this by running multiple realizations to assess statistical robustness.}

\tr{The most counterintuitive result may be that doubling the initial gas mass does not lead to a more massive NSC.
While the extra gas does temporarily enhance star formation at the very start of the simulation, this does not translate into long-term NSC growth.
One plausible mechanism is that the initial burst of star formation---stimulated by the extra gas---triggers vigorous feedback that expels much of the gas reservoir.
This self-regulating loop, consistent with simulations and analytic models, prevents sustained gas accretion onto the NSC despite the higher initial gas mass \citep{2017MNRAS.465.1682H,2025OJAp....8E...7T}.
It is important to note that our simulation with twice the gas mass does not settle into a stable galaxy with double the gas; the initial conditions therefore do not map directly onto a stabilized galaxy.}

\tr{The rotational properties of NSCs (Sec.~\ref{sec:result_rotation}) provide additional evidence for these degeneracies.
The increase in stellar angular momentum is generally consistent with gas accretion followed by in-situ star formation and coplanar mergers \citep{2019A&A...629A..44L,fahrion2019A&A...628A..92F}.
However, distinguishing between angular momentum gained through gas inflow and that gained through coplanar mergers remains difficult.
All four runs show broadly consistent evolution toward higher ellipticity and specific angular momentum, with substantial overlap with observational data from \citet{2019A&A...629A..44L} and \citet{pinna2021ApJ...921....8P}.
These trends support the idea that rotation is imprinted during assembly through both gas inflow and cluster merging processes \citep{2008ApJ...687..997S,2017MNRAS.464.3720T}, but they also highlight the inherent degeneracies in using kinematics alone to distinguish formation channels.}

\tr{In summary, we observe numerous degeneracies among physical processes.
NSC growth appears to be self-regulating: mergers can suppress in-situ star formation by disrupting core gas, while feedback-regulated gas dynamics can compensate for fewer mergers by enhancing in-situ star formation---provided sufficient gas is available.
Recent studies have explored this interplay between accretion of globular clusters and in-situ star formation in NSC mass buildup \citep[e.g.,][]{2022A&A...658A.172F}.
These intertwined behaviors make it difficult to cleanly isolate the individual contributions of each mechanism.
Additional caveats and limitations are discussed in Sec.~\ref{sec:discussion_limitations}.}


\subsection{Numerical Validation}
\label{sec:discussion_numerical}

Two numerical aspects of our simulations---core collapse dynamics and dark matter resolution---require validation to ensure they do not compromise the astrophysical results reported above.
We present the full convergence tests in Appendix~\ref{sec:appendix_numerical} and summarize the key conclusions here.

The artificially short core-collapse timescale ($\sim40\Myr$), which arises because our star particles have masses of $300\msun$ rather than representing individual stars, is the primary constraint on our simulation runtime.
Comparison of our NSC embedded in a live galaxy with isolated star cluster simulations run in both \nbody{} and {\tt Nbody6++GPU} confirms that: (i)~core collapse is delayed in the galactic environment due to dynamical heating from the background potential, and (ii)~physical processes such as mergers and star formation dominate the evolution of the NSC over purely collisional relaxation effects (Fig.~\ref{fig:core_collapse}).
In a realistic star-by-star treatment with $\sim10^6$ stars, the core-collapse time would extend to $\sim\mathrm{Gyr}$, removing this constraint on runtime.

Regarding dark matter, our NSCs accrete dark matter to masses comparable to or exceeding their stellar mass after $\sim80\Myr$, regardless of the gravitational softening length or particle mass resolution (Fig.~\ref{fig:dark_matter}).
However, the impact on internal cluster dynamics is confined primarily to the outer regions (beyond the 30\% Lagrangian radius), while the inner structure---where star formation and merger-driven evolution occur---remains converged across resolutions (Fig.~\ref{fig:core-collapse_dm}).
The accretion of dark matter may be linked to the core--cusp problem \citep[e.g.,][]{herlan2023MNRAS.523.2721H}, and could in principle depend on the dark matter model; however, it does not influence the primary results of this work.


\subsection{Limitations \& Caveats}
\label{sec:discussion_limitations}

\tr{Our simulations are subject to several important caveats. First, the realism of star formation and stellar feedback depends heavily on our modeling choices, which operate at two distinct levels.
At the level of the {\it feedback scheme}, dwarf galaxies are particularly sensitive: different supernova feedback implementations (e.g., superbubble vs.\ blastwave) produce qualitatively different star formation histories and outflow properties \citep{2024ApJ...970...40A}, because bursty star formation and feedback-driven outflows are hallmarks of dwarf galaxy evolution.
At the level of {\it parameter tuning within a given scheme}, however, dwarf galaxies are comparatively less sensitive than massive galaxies, where the delicate interplay between AGN feedback, cooling flows, and star formation creates a more demanding calibration problem \citep{2012MNRAS.426..140D}.
Our own experience illustrates this resolution dependence concretely: our simulations adopt the fiducial subgrid parameter set originally tuned by \citet{kim2011ApJ...738...54K}, but the 32-times finer spatial resolution of this work ($0.625\pc$) demanded re-calibration \citep{2015MNRAS.450.1937C,2018MNRAS.475..648P,2018MNRAS.477.1578H}.
Higher resolution increases the effectiveness of both thermal and kinetic stellar feedback \citep{2012MNRAS.426..140D,2018MNRAS.478..302S}; when we re-ran the simulations with the original parameters at our new resolution, the host galaxy became dissipated owing to overly vigorous supernova heating.
To mitigate this, we lowered the mass-to-energy conversion factor for supernova ejecta by an order of magnitude (see Sec.~\ref{sec:method_star_physics}); no further calibration was performed.
This sensitivity to resolution underscores that our results may shift under alternate feedback implementations or calibration strategies.}

\tr{Second, we do not perform star-by-star simulations. 
Although the treatment of collisional dynamics within star clusters remains fully intact, our star particles do not represent single stars in this work.
True star-by-star simulations require modeling the formation of each star along with detailed astrophysics such as supernova and wind feedback, supernova remnants, and more.
We are currently implementing a star-by-star framework equipped with radiation transfer and ten chemical species, developed by the AEOS simulations, into \newenzo{}.
With this in place, we anticipate producing significantly more sophisticated results in future studies.}

\tr{Third, the simulation runtime of 150~Myr captures only an early growth phase of the NSC.
The primary constraint on the runtime is the artificially short core-collapse timescale of $\sim40\Myr$, which arises because our star particles have masses of $300\msun$ rather than representing individual stars (see Appendix~\ref{sec:appendix_core_collapse}).
After core collapse, the internal dynamics of the NSC become unreliable without proper binary heating, making extended evolution beyond $\sim3$--$4$ core-collapse times physically questionable.
Nevertheless, 150~Myr is sufficient to capture the processes most relevant to this study for three reasons.
First, dynamical friction timescales depend strongly on orbital radius: for the $\sim10^5\msun$ star clusters that form within the direct N-body region ($r<200\pc$), the inspiral timescale is only $\sim20$--$40\Myr$ (Sec.~\ref{sec:result_merger}), well sampled by our runtime.
Clusters forming at larger radii ($r>1\kpc$) require $\sim\mathrm{Gyr}$ to arrive and would not contribute within the timeframe set by our particle resolution constraints.
Second, the key processes of in-situ star formation---gas cooling, inflow, and collapse within the NSC---operate on timescales of a few to tens of Myr, all well resolved.
Third, we do not claim to model the full NSC lifetime; rather, our simulations are designed to study the interplay between mergers and in-situ star formation and to identify the self-regulatory behavior we report.
Extending to longer timescales---ideally with star-by-star resolution to avoid the core-collapse limitation---would be necessary to assess whether these trends persist over a full NSC lifetime.}

\tr{Fourth, our initial conditions introduce an artificial rapid gas inflow. 
We initialize gas rotation to match the host galaxy rather than the NSC, creating a mismatch in angular momentum. 
This leads to a dramatic early infall of gas into the NSC region, which may inflate early growth rates. 
A more carefully matched initial rotation curve would mitigate this artifact.}

\tr{Fifth, we lack sufficient resolution to model gas and dark matter inside individual star clusters; however, gravitational dynamics of stars within star clusters are resolved robustly. 
Limited force softening and particle discreteness artificially accelerate core collapse or disruption in small-$N$ clusters---a phenomenon documented in dark matter subhalo simulations and likely applicable to our stellar clusters \citep{2018MNRAS.474.3043V}. 
This artificial disruption occurs early in the run and may skew merger rates and mass growth, raising concerns about realism.}

\tr{To draw firmer conclusions, we need additional simulations that simultaneously vary subgrid parameters to reduce stochasticity, extend to longer timescales, improve spatial resolution, and refine initial angular momentum configurations. 
These steps are essential to isolate the roles of mergers, in-situ star formation, feedback, and gas dynamics in NSC evolution.}

\section{Summary}
\label{sec:summary}

In this study, we introduce a simulation framework that enables the investigation of the evolution of star clusters within a live galaxy, along with star formation, supernova feedback, and collisional gravity.
With this, we investigate the evolution of Nuclear Star Clusters (NSCs) in dwarf galaxies, focusing on the relative contributions of two key mechanisms: mergers of star clusters and in-situ star formation.
Our simulation framework is based on \oldenzo{} \citep{2024ApJ...974..193J} and further developed to integrate a new in-house direct N-body code, \nbody{} (see Sec.~\ref{sec:method_ac-scheme}; Jo et al., in prep.), replacing the previously utilized {\tt Nbody6++GPU} as the direct N-body solver, which is extensively compared to the simulation of individual star clusters.
\nbody{} is engineered to handle complex and adaptive stellar systems within galaxies, while simultaneously supporting the execution of subgrid physics processes for star particles, such as star formation and supernova feedback that interact with the hydrodynamics. 
With this, it is possible to simulate both the host galaxy and the nuclear star cluster (NSC) simultaneously, incorporating live dark matter, gas, and stars within \newenzo{}, while resolving star cluster internal dynamics at a level comparable to that achieved by the direct N-body community.

Using \newenzo{}, we model a dwarf galaxy with a total mass of $\sim 10^{10}\msun$ serving as our host. 
We initially place a $3\times10^5\msun$ NSC, following the Plummer profile with an effective radius of $2 \pc$, at the galaxy's center in the initial conditions of the simulation. 
We focus on how this NSC evolves through mergers and in-situ star formation, and examine how stellar physics parameters---star formation efficiency and supernova feedback---affect NSC evolution by performing three additional simulations that vary these parameters, resulting in four simulations in total: `Fiducial', `No Feedback', `Low Star Formation', and `$2\times$\,Gas'.

Our simulations span $150 \myr$, capturing only an early phase of NSC evolution, due to computational constraints (see Sec.~\ref{sec:discussion_limitations}).
Starting from idealized dwarf galaxy initial conditions, two processes primarily drive NSC growth.
First, gas fragmentation promotes the formation of gas clumps and subsequently star clusters (see Sec.~\ref{sec:result_overview}).
Some of these clusters migrate toward the central regions via dynamical friction and tidal forces, eventually merging with the NSC (see Sec.~\ref{sec:result_merger}). 
Second, surrounding gas is drawn into the NSC, fueling in-situ star formation (see Sec.~\ref{sec:result_insitu}).

Mergers significantly contribute to NSC mass growth, especially during major merger events that produce pronounced mass jumps (Sec.~\ref{sec:result_merger}). 
We find some indications that the number of mergers depends on the choice of stellar physics---such as star formation efficiency and the strength of supernova feedback---as well as the gas content of the galaxy.
However, the differences in merger frequency across runs are modest (see Fig.~\ref{fig:merger_tree}), whereas the mass growth contributed by mergers varies more notably (see Fig.~\ref{fig:mass_growth_merger}).
Despite these variations, the final NSC masses at 150~Myr are comparable across all four simulations.

Gas inflows toward the galactic center enable in-situ star formation within the NSC, while supernova feedback regulates this process. 
Our simulations indicate that the role of supernova feedback within NSCs is highly degenerate: hot, feedback-driven gas contributes negligibly to the overall NSC mass and star formation rates (see Fig.~\ref{fig:mass_growth_insitu}). 
Although supernova feedback temporarily heats NSC gas and intermittently suppresses star formation, gravitational forces generally dominate due to the high densities involved, making feedback's sustained impact marginal (Sec.~\ref{sec:result_insitu}). 
The gas mass within NSCs remains relatively constant at $\sim10$--$20\%$ of their total stellar mass, aside from initial inflows driven by angular momentum imbalances (see Figs.~\ref{fig:gas_properties} and \ref{fig:angular_momentum}).

In the `Fiducial' model and `$2\times$\,Gas' case, mergers and in-situ star formation contribute approximately equally to the NSC mass growth, with mergers slightly dominating. 
In the `No Feedback' and `Low Star Formation' scenarios, mergers clearly dominate, accounting for more than 50\% of the mass increase, whereas in-situ star formation contributes less than 20\%. 
The remaining $\sim 30\%$ is composed of stars initially present or randomly accreted from the surrounding environment. 
A complex tension remains between the relative roles of in-situ star formation and mergers due to inherent degeneracies---for instance, mergers can regulate subsequent in-situ star formation by disrupting the dense core gas reservoir (Sec.~\ref{sec:discussion_merger_regulates}). 
In addition, the current limitations---such as the lack of individual star-by-star physics and limited spatial and particle-mass resolutions---prevent definitive conclusions at this stage. 
However, future improvements (see Sec.~\ref{sec:future_work}) are expected to allow our simulation framework to serve as a cornerstone for disentangling the intricate interplay between mergers and in-situ star formation, ultimately advancing our understanding of the formation and evolution of NSCs.

\section{Future Work}
\label{sec:future_work}
This work lacks two important features---individual star physics and black holes. 
In the future work, we will address the following:
\begin{itemize}
    \item We will implement a star-by-star recipe to resolve more accurate evolutionary pathways for the NSCs. 
    In addition, we are planning on applying the direct N-body to the entire galaxy so that we can trace the evolution of the NSCs more accurately.
    Furthermore, we will run a number of simulations with different initial conditions for host galaxies to investigate formation pathways for NSCs.
    
    \item It is widely believed that NSCs harbor intermediate-mass black holes and are accompanied by SMBHs. 
    Since \nbody{} can seamlessly work with \enzo{}'s subgrid physics, we anticipate simulating NSCs with black holes which are subject to black hole accretion and black feedback such as jets.
\end{itemize}

\acknowledgments
The Flatiron Institute is supported by the Simons Foundation.
We thank Eunwoo Chung at Seoul National University for running {\tt Nbody6++GPU} for comparison in Sec. \ref{sec:appendix_core_collapse}.
Ji-hoon Kim’s work was supported by the National Research Foundation of Korea (NRF) grant funded by the Korea government (MSIT) (No. 2022M3K3A1093827 and No. 2023R1A2C1003244).  His work was also supported by the Global-LAMP Program of the NRF grant funded by the Ministry of Education (No. RS-2023-00301976).  His work was also supported by the National Institute of Supercomputing and Network/Korea Institute of Science and Technology Information with supercomputing resources including technical support, grants KSC-2022-CRE-0355 and KSC-2024-CRE-0232.


\appendix

\section{Numerical Convergence Tests}
\label{sec:appendix_numerical}

This appendix presents the convergence tests summarized in Sec.~\ref{sec:discussion_numerical}.
We assess two numerical aspects---core collapse dynamics and dark matter resolution---that validate the robustness of the astrophysical results reported in the main text.

\subsection{Core Collapse and Expansion}
\label{sec:appendix_core_collapse}

\begin{figure}
    \centering
    \includegraphics[width=0.5\linewidth]{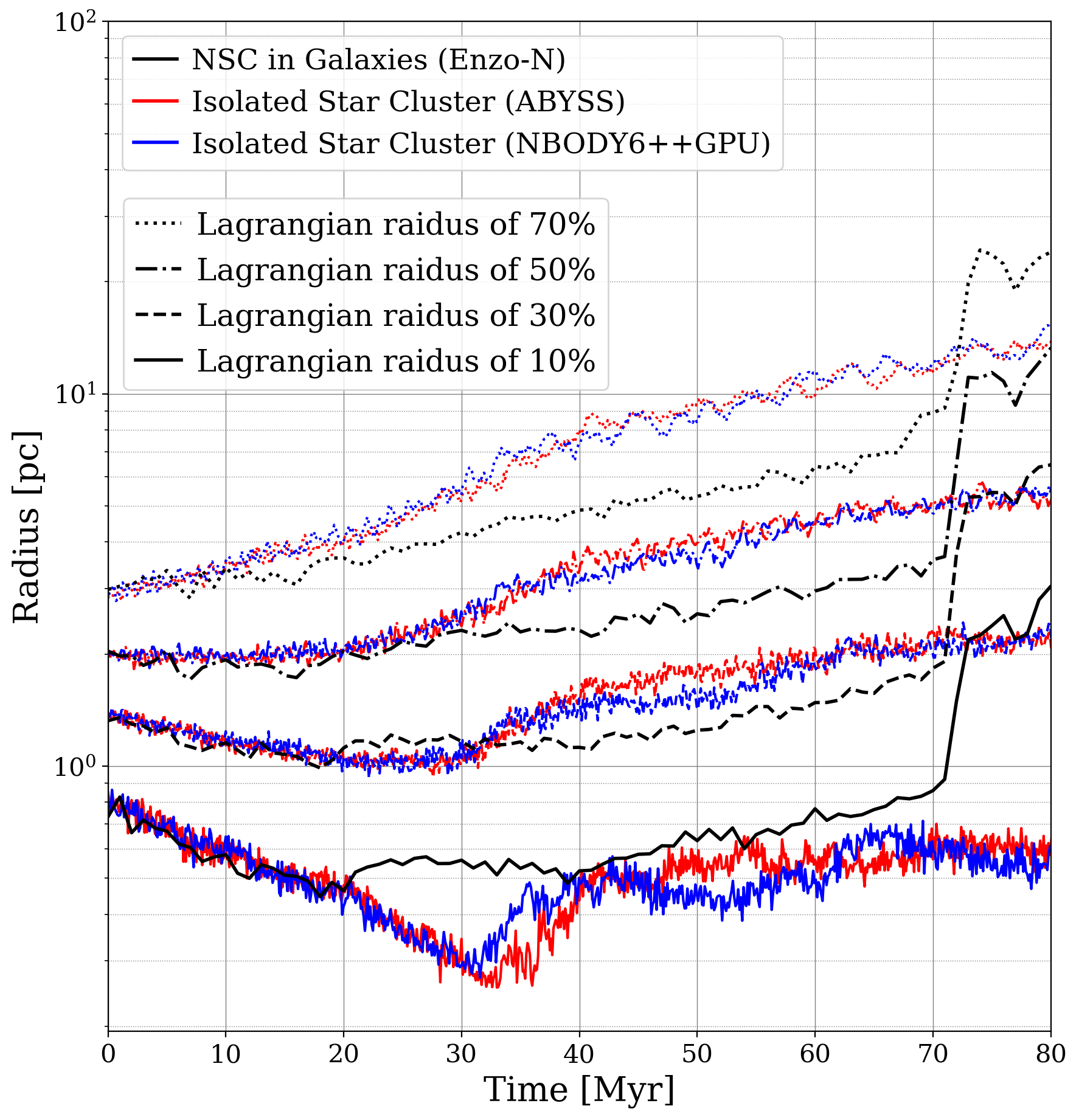}
    \caption{Lagrangian radii of the star clusters over 80 Myr.
    The evolution of the isolated star clusters are tested in \nbody{} ({\it red}) and {\tt Nbody6++GPU} ({\it blue}) codes, while the evolution of the NSC is run with \newenzo{} ({\it black}) within a dwarf galaxy without star formation and supernova feedback.
    Four different Lagrangian radii are represented by {\it dotted} for 70\%, {\it dot-dash} for 50\%, {\it dashed} 30\%, and {\it solid} for 10\%.
    The Lagrangian radius is a radius that encompasses a given fraction of mass of the star clusters.
    }
    \label{fig:core_collapse}
\end{figure}

Core collapse and expansion are crucial phases in the life cycle of star clusters, influencing their evolution and the fate of their constituent stars.
Expansion of star clusters consists of two conceptually distinct mechanisms---ejection and evaporation.
An encounter with a star in close proximity can lead to a velocity change similar to the original velocities of the stars involved, potentially resulting in one star gaining a velocity greater than the local escape velocity; this phenomenon is known as ejection \citep[][]{binney1987gady.book.....B}. 
Conversely, evaporation arises from a succession of less intense and more remote interactions that can incrementally boost a star's energy until a final minor encounter grants it just enough positive energy to escape.
On the other hand, as the outer part of the cluster expands, the energy loss results in an increase in kinetic energy of the remaining stars along with a decrease in the potential energy due to the virial theorem.
This process, known as core collapse, leads to a contraction of the system and kinetic energy increases of the particles, resulting in a dramatic growth in the central density.

\tr{To investigate the impact of the live galaxy on core collapse, we run three additional simulations: one NSC within a live galaxy without star formation and supernova feedback, and two isolated star cluster simulations without a galaxy using {\tt Nbody6++GPU} and \nbody{}.}
Fig.~\ref{fig:core_collapse} illustrates the comparison of the Lagrangian radii between the isolated star clusters and the NSC within a dwarf galaxy.
\tr{The Lagrangian radius is a radius that encompasses a given fraction of mass of the star clusters.}
The NSC in a dwarf galaxy is run with \newenzo{} ({\it black}) without stellar physics, while the isolated star clusters are run with \nbody{} ({\it red}) and {\tt Nbody6++GPU} ({\it blue}).
We present four different Lagrangian radii: {\it dotted} for 70\%, {\it dot-dash} for 50\%, {\it dashed} 30\%, and {\it solid} for 10\%.
First, the evolution of the isolated star clusters is consistent between \nbody{} and {\tt Nbody6++GPU}, exhibiting core collapse at around 30~Myr.
However, in the case of the NSC in the galaxy, the core collapse halts at $\sim 20$~Myr.
The expansion of the NSC is also slower than that of the isolated star clusters.
This can be attributed to the tidal influence of gas and dark matter dynamics.
For instance, the core collapse can be suppressed due to dynamical heating from the background potential---gas and dark matter.
In addition, gas infall costs extra energy from the NSCs.
The sudden change at 70~Myr arises from a tidal interaction with a nearby gas clump.

For the isolated star clusters, theoretical predictions yield that core collapse occurs at $\tau_\mathrm{cc}\approx200$--$300\,t_\mathrm{r}(r=0)$ or $\tau_\mathrm{cc} \approx 20 \,t_\mathrm{rh}$, where $\tau_\mathrm{cc}$, $t_\mathrm{r}(r=0)$, and $t_\mathrm{rh}$ are the core-collapse time, relaxation time of the core, and half-mass relaxation time.
The core relaxation time can be written as 
\begin{equation}
    t_\mathrm{r}(r=0) = \frac{0.065 v^3_m}{nm^2G^2\ln\Lambda},
\end{equation}
where the Coulomb logarithm is
\begin{equation}
    \ln \Lambda \equiv \ln(p_\mathrm{max}/p_0)\approx \ln(0.4N).
\end{equation}
\tr{Here, $G$, $n$, $m$, and $v_m$ are the gravitational constant, the number density at the center, the mass of stars, and the 1D velocity dispersion.}
For our star cluster with mass of $3\times10^5\msun$ and $10^3$ particles, $t_\mathrm{r}(r=0) \sim 10^5\yr$ at $0\myr$.
This gives an approximate core collapse time of $20$--$30\myr$ using $\tau_\mathrm{cc}=200$--$300\,t_\mathrm{r}(r=0)$, consistent with the isolated simulation results.

However, in a more realistic scenario where individual stars are considered, the core-collapse time is expected to be $\sim\mathrm{Gyr}$ due to the presence of a greater number of particles, approximately $10^6$ stars.
This disparity stems from variations in the relaxation time, which lead to both accelerated expansion and contraction of our star clusters compared to the realistic case.
Our analysis indicates that \tr{physical processes such as mergers and star formation} are the predominant factors in the evolution of the NSCs compared to expansion and core collapse.
Nonetheless, employing a star-by-star methodology is anticipated to mitigate mass loss via expansion and result in a delayed core collapse.

\subsection{Dark Matter Resolution Effects}
\label{sec:appendix_dark_matter}

\begin{figure}
    \centering
    \includegraphics[width=0.5\linewidth]{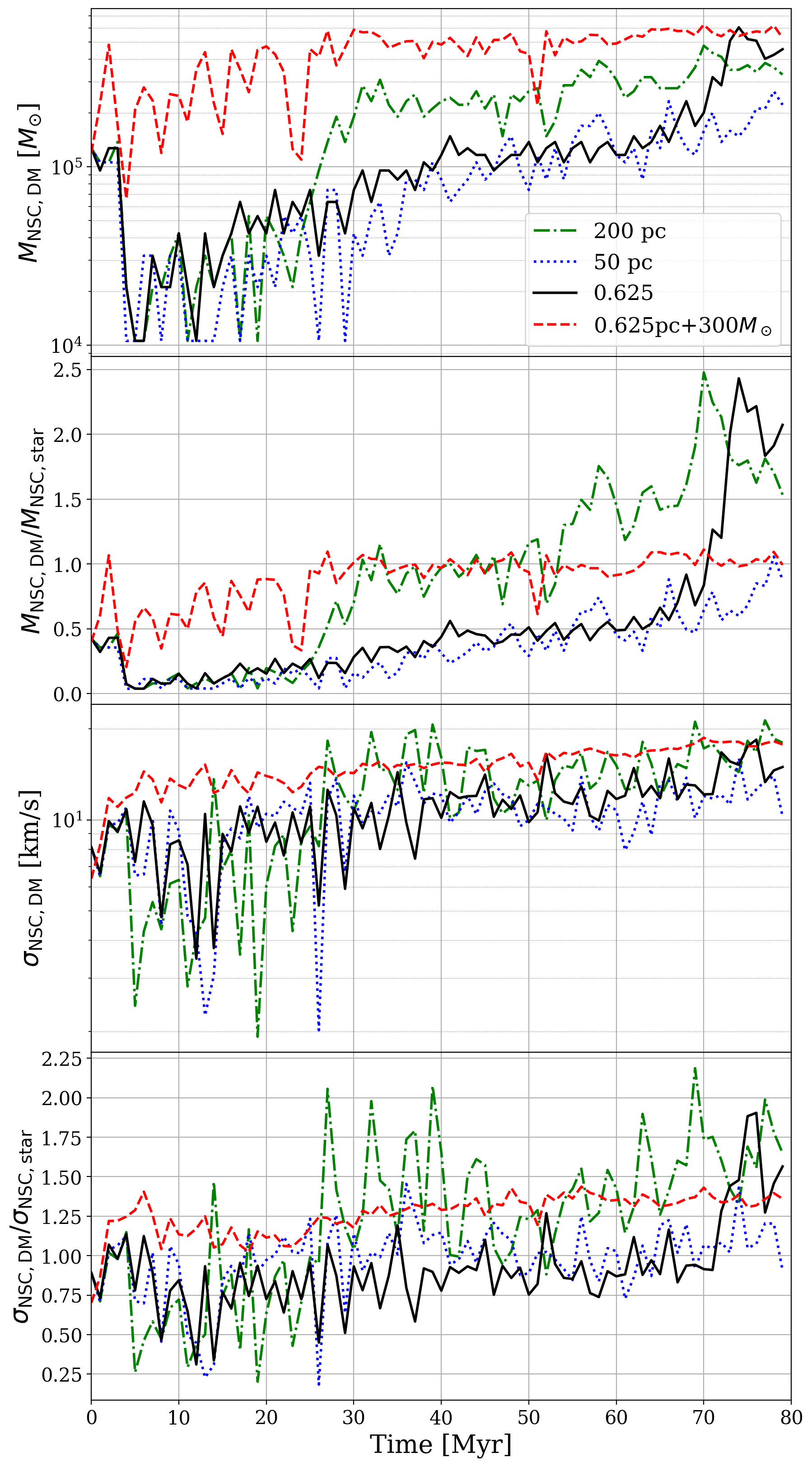}
    \caption{The evolution of dark matter mass, ratio of dark matter mass to the NSC mass, velocity dispersion, and the ratio of velocity dispersion within the NSCs from the top, with respect to gravitational softening lengths and mass resolution.
    The resolution of the simulations include $200 \pc$ ({\it green dot-dash}), $50\pc$ ({\it blue dotted}), $0.625\pc$ ({\it black solid}) with the dark matter particle mass of $10^4\msun$, and $0.625 \pc$ with the dark matter particle mass of $300\msun$ ({\it red dashed}).
    The spatial resolution in this context denotes the smallest cell size at which the particle-mesh solver calculates gravity for the particles.
    Notice that $0.625\pc$ with $10^4\msun$ is the fiducial resolution for this work.}
    \label{fig:dark_matter}
\end{figure}

\begin{figure}
    \centering
    \includegraphics[width=0.5\linewidth]{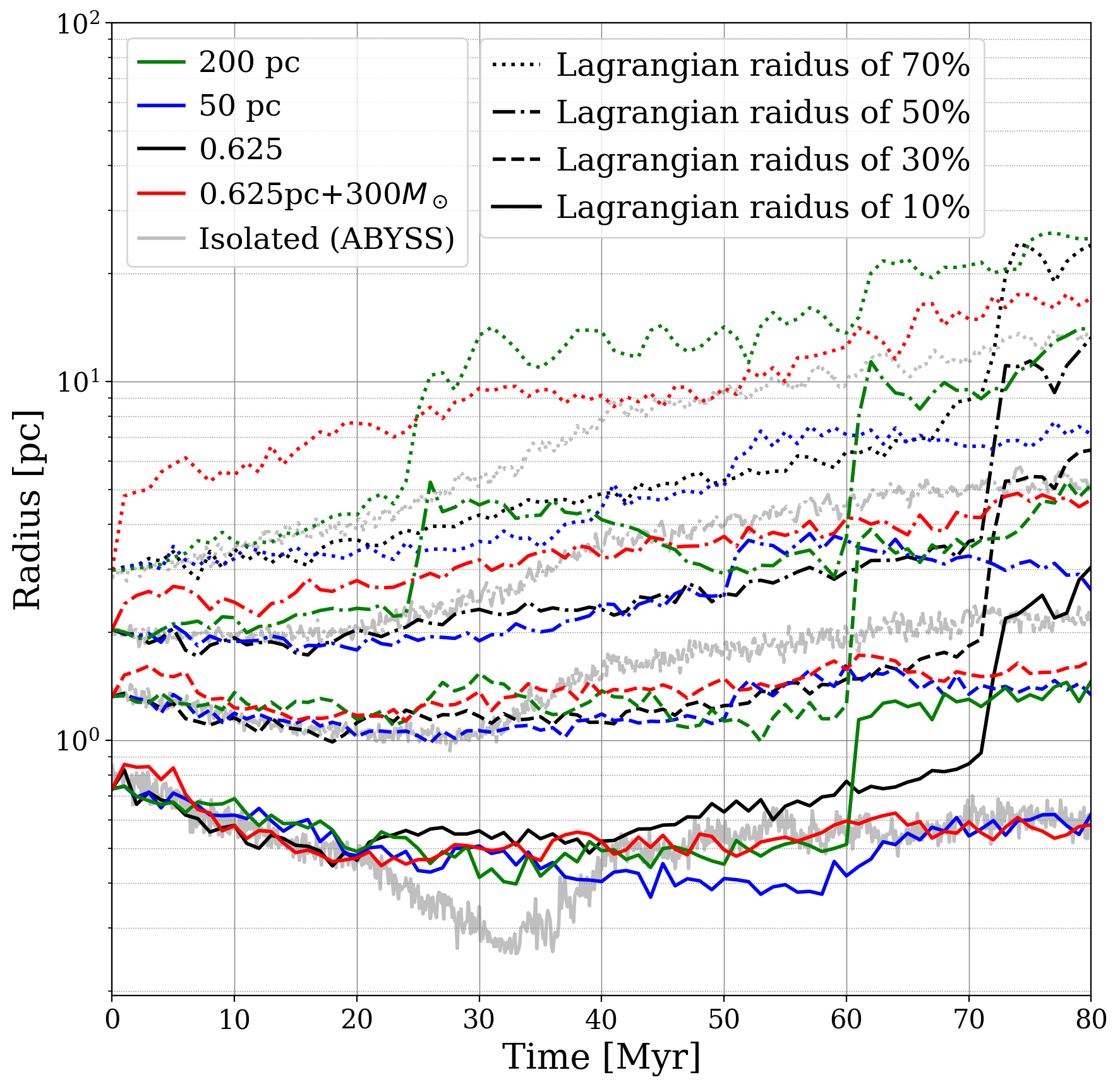}
    \caption{Effect of dark matter gravitational softening lengths and mass resolution in evolution of the NSCs within the host galaxies without star formation and supernova feedback.
       The resolution of the simulations include $200 \pc$ ({\it green}), $50\pc$ ({\it blue}), $0.625\pc$ ({\it black}) with the dark matter particle mass of $10^4\msun$, and $0.625 \pc$ with the dark matter particle mass of $300\msun$ ({\it red}).
       The isolated star cluster, run with \nbody{}, is presented by the {\it grey} lines.
       Four different Lagrangian radii are represented by {\it dotted} for 70\%, {\it dot-dash} for 50\%, {\it dashed} 30\%, and {\it solid} for 10\%.}
    \label{fig:core-collapse_dm}
\end{figure}

One noteworthy observation is the dark matter distribution within \tr{the radius of the NSCs}.
In the initial conditions, dark matter does not exist in the central region where the NSC resides due to the particle mass resolution of $10^4\msun$ (refer to the {\it green dotted} lines in Fig.~\ref{fig:profile}). 
After several Myr, dark matter particles fill up the NSC following its profile, while the dark matter profile outside the NSC remains little changed.

However, due to the mass resolution of the dark matter particles, there exist only $\sim 10$ dark matter particles within $10\pc$ and six within $1\pc$, leading to significant statistical uncertainty and \tr{errors} due to discreteness.
Regarding the resolution of dark matter in hydrodynamic simulations, \citet{regan2015MNRAS.449.3766R} have studied the impact of dark matter particle resolution on \tr{the collapse of gas} using \enzo{}.
They find that converged results require $M_\mathrm{core}/M_\mathrm{DM} > 100$, where $M_\mathrm{core}$ is the enclosed baryon mass within the core and $M_\mathrm{DM}$ is the minimum dark matter particle mass.
Otherwise, artificial tidal heating can occur.

We run four additional simulations with varying softening lengths to investigate these resolution effects.
As in Sec.~\ref{sec:appendix_core_collapse}, the simulations are run without stellar physics to isolate the resolution effects.
The smoothing lengths of the dark matter particles are controlled by the {\it MaximumParticleRefinementLevel} in \enzo{}.
We choose three smoothing lengths---$200\pc$, $50\pc$, and $0.625\pc$---with a dark matter particle mass of $10^4\msun$, where $0.625\pc$ is the finest cell size of the simulations.
Additionally, we vary the mass of the dark matter from $10^4\msun$ to $300\msun$ with a constant smoothing length of $0.625\pc$.

Fig.~\ref{fig:dark_matter} shows (top to bottom) the evolution of the dark matter mass, the ratio of dark matter mass to NSC mass, the velocity dispersion, and the ratio of velocity dispersion within the NSCs, for a range of gravitational softening lengths and mass resolutions.
\tr{Here, dark matter particles are defined as those located within the NSC radius determined by \texttt{HDBSCAN}.}
One important result is that, regardless of resolution, the NSCs attain dark matter masses comparable to or exceeding their stellar mass after 80~Myr, which can potentially affect the evolution and properties of NSCs.
A consistent pattern across different smoothing lengths is not clearly discernible; the fluctuations can largely be attributed to noise.
On the other hand, the higher mass resolution ({\it red}) evidently results in a more stable dark matter distribution, with less fluctuation in both mass and velocity dispersion.

Fig.~\ref{fig:core-collapse_dm} illustrates the Lagrangian radii with different resolutions to investigate the impact on the internal dynamics of star clusters.
The variation in smoothing lengths does not significantly affect the evolution of the NSCs.
The occasional jumps are attributed to collisions and tidal interactions with gas clumps.
The impact of mass resolution is predominantly observed in the outer regions of the NSCs: the high mass resolution run ({\it red}) leads to an overall expansion, with the 70\% Lagrangian radius significantly larger than in the other runs, while the 10\% and 30\% radii converge.

In conclusion, NSCs can accrete dark matter to some extent \citep[cf.][]{herlan2023MNRAS.523.2721H,balaji2023JCAP...08..063B}, and this dark matter can influence the evolution of NSCs, mostly in the outer regions.
The accretion of dark matter may be linked to the core--cusp problem of dark matter profiles within halos \citep[e.g.,][]{herlan2023MNRAS.523.2721H}, and the results could in principle depend on the dark matter model (e.g., cold, warm, or self-interacting).
Nevertheless, as the main focus of this work lies in the impact of mergers and in-situ star formation, the dark matter physics is not expected to influence the primary results.

The challenges associated with numerically simulating dark matter---discreteness and resolution effects---are widely documented \citep{steinmetz1997MNRAS.288..545S,melott1997ApJ...479L..79M,dolag2009MNRAS.399..497D,christensen2010ApJ...717..121C,vazza2011MNRAS.418..960V,regan2015MNRAS.449.3766R,power2016MNRAS.462..474P, hopkins2023MNRAS.525.5951H,liu2023MNRAS.522.3631L}.
For particle-mesh methods, robustness requires more than one particle per softening length, and the smoothing length should be comparable to the mean inter-particle separation \citep{melott1997ApJ...479L..79M}.
Meeting these conditions is practically impossible in adaptive mesh refinement codes where refinement is driven by gas dynamics at small scales.
It is therefore important to acknowledge that spurious gravitational effects may be present at the smallest scales.
A definitive assessment of which resolution yields physically accurate results is left to future work.

\bibliography{main}{}
\bibliographystyle{aasjournal}

\end{document}